\documentclass[a4paper,11pt]{article}
\pdfoutput=1 

\usepackage{jheppub,subcaption,xspace,graphicx,todonotes,cleveref,soul} 

\usepackage[normalem]{ulem}
\usepackage[utf8]{inputenc}

\newcommand{\HEJ}{{\tt HEJ}\xspace}
\newcommand{\ariadne}{A\protect\scalebox{0.8}{RIADNE}\xspace}
\newcommand{\as}{\ensuremath{\alpha_s}\xspace}
\newcommand{\HIGHEJ}{\emph{High Energy Jets}\xspace}
\newcommand{\py}{P\protect\scalebox{0.8}{YTHIA}8\xspace}
\newcommand{\pyt}{P\protect\scalebox{0.8}{YTHIA}\xspace}
\newcommand{\pysix}{P\protect\scalebox{0.8}{YTHIA}6\xspace}
\newcommand{\madgraph}{M\protect\scalebox{0.8}{AD}G\protect\scalebox{0.8}{RAPH}5\_aMC@NLO\xspace}
\newcommand{\MeV}{\ifmmode {\mathrm{\ Me\kern -0.1em V}}\else
  \textrm{~Me\kern -0.1em V\xspace}\fi}%
\newcommand{\GeV}{\ifmmode {\mathrm{\ Ge\kern -0.1em V}}\else
  \textrm{~Ge\kern -0.1em V\xspace}\fi}%
\newcommand{\TeV}{\ifmmode {\mathrm{\ Te\kern -0.1em V}}\else
  \textrm{~Te\kern -0.1em V\xspace}\fi}%

\newcommand{\Ca}{\ensuremath{C_{\!A}}\xspace}

\newcommand{\Nc}{\ensuremath{N_{\!C}}\xspace}
\newcommand{\gs}{\ensuremath{g}}

\newcommand{\eg}{\emph{e.g.\xspace}}
\newcommand{\HEJpy}{{\tt HEJ}+\pyt}


\subheader{
  \begin{flushright}\sffamily
   CPT/17/184\  IPPP/17/92\\
    LU-TP 17-38\\
    MCnet-17-22\\
    CoEPP-MN-17-22
  \end{flushright}
}

\title{\boldmath Merging High Energy with Soft and Collinear Logarithms using HEJ
  and PYTHIA}

\author[a]{Jeppe R.~Andersen,}
\author[a,b]{Helen M.~Brooks}
\author[c]{and Leif L\"onnblad}

\affiliation[a]{Institute for Particle Physics Phenomenology, University of Durham,
  South Road, Durham DH1 3LE, UK}
\affiliation[b]{School of Physics and Astronomy, Monash University, Clayton, VIC 3800, Australia}
\affiliation[c]{Dept.~of Astronomy and Theoretical Physics, Lund University,
  Sweden}

\emailAdd{jeppe.andersen@durham.ac.uk}
\emailAdd{helen.brooks@monash.edu}
\emailAdd{Leif.Lonnblad@thep.lu.se}

\abstract{We present a method to combine the all-order treatment of the
  \HIGHEJ exclusive partonic Monte Carlo (\HEJ) with the parton shower of \py, while
  retaining the logarithmic accuracy of both. This procedure enables 
  the generation of fully realistic and hadronised events with \HEJ. 
  Furthermore, the
  combination of the two all-order treatments leads to improvements in the
  quality of the description of observables, in particular for regions with
  disparate transverse scales.
}

\begin{document} 
\maketitle
\flushbottom

\section{Introduction}
\label{sec:intro}
The analyses of collider data collected at both the Tevatron \cite{Abazov:2013gpa}
($\sqrt{s}=1.96$\TeV) and the LHC \cite{Aad:2011jz,Aad:2014pua,Aad:2014qxa,Chatrchyan:2012gwa,Aad:2015nda,Khachatryan:2016udy}
($\sqrt{s}=7,8$\TeV)
indicate that perturbative terms beyond fixed order are required for the
description of observables in processes involving at least two jets, in the
region of large partonic centre-of-mass energy $\sqrt{\hat s}$ compared to
the typical transverse momentum scale $p_\perp$,
$\sqrt{\hat s}/p_\perp>5$. This corresponds to the jets spanning
more than 3 units of rapidity. It is of course well-known
and indeed not surprising that the convergence of the perturbative series
requires input beyond fixed-order perturbation theory in certain regions of
phase space. The dominant and large corrections in this particular region of
phase space is the focus of one of the applications of the theory of
Balitsky-Fadin-Kuraev-Lipatov
(BFKL) \cite{Fadin:1975cb,Kuraev:1976ge,Kuraev:1977fs,Balitsky:1978ic}. The
study of such perturbative effects has received renewed interest, not only because the
increasing energy of colliders allows for detailed study of
observables in these regions, but also because some 
measurements specifically concentrate on experimental signatures with 
large rapidity spans \cite{Chatrchyan:2012pb,Chatrchyan:2012gwa,Chatrchyan:2013jya,Aad:2014pua,Aad:2014qxa,Aad:2015nda,Khachatryan:2016udy,Aaboud:2017fye}.

The BFKL formalism can provide a systematic description of large
perturbative corrections in two separate kinematic 
limits. Firstly, in the small-$x$ limit of $\hat{s}/s\ll 1$ (where $\sqrt{\hat{s}}$ is
the centre-of-mass energy for the partonic process and $\sqrt{s}$ is the
centre-of-mass energy of the collider), BFKL theory can be used to
describe the evolution of the PDFs in $x$. Secondly, in the limit $\hat s/|\hat t|\gg 1$
(where $|\hat t|^{1/2}$ is a typical jet transverse momentum scale and
$\hat s\le s$), BFKL theory captures the single-logarithmic
corrections in $\hat s/|\hat t| \sim e^{\Delta y}$ to the hard-scattering matrix element for
processes with a colour-octet exchange between two jets. These two applications
of BFKL are valid in two opposite kinematic regions. 
In the former case, a formalism to combine logarithms of BFKL and
Dokshitzer-Gribov-Lipatov-Altarelli-Parisi
(DGLAP) \cite{Dokshitzer:1977sg,Gribov:1972ri,Altarelli:1977zs} origin
was developed resulting in the Ciafaloni-Catani-Fiorani-Marchesini
(CCFM) equation \cite{Ciafaloni:1987ur,Catani:1989yc,Catani:1989sg,Marchesini:1994wr},
with an explicit partonic evolution implemented in
\textsc{Cascade} \cite{Jung:2000hk,Jung:2010si}. This has the non-perturbative stages of
the evolution handled by \pysix\cite{Sjostrand:2006za}.

The second \textit{high energy} limit of $\hat s/|\hat t|\gg 1$ is the focus of
Mueller-Navelet-style studies of QCD processes involving at
least two jets \cite{Mueller:1986ey}. Monte Carlo approaches to solving the BFKL equation were developed for the detailed study of
such processes \cite{Orr:1997im,Orr:1998ps}. A more accurate description of the scattering matrix
elements that still captures the BFKL logarithms was later obtained, which also 
included matching to fixed-order high multiplicity matrix
elements \cite{Andersen:2009nu,Andersen:2009he,Andersen:2011hs}. 
This approach
is implemented in the parton-level Monte Carlo of \HIGHEJ (\HEJ), 
which shall be described further in \cref{sec:hej}. 

An application of the \HEJ formalism that is of particular interest is the study
of the production of a Higgs boson in association with dijets that have a
large rapidity separation. 
The ability to model jets in the rapidity interval
permits an examination of the sensitivity of predictions
to the placement of vetoes upon additional radiation, 
which is particularly relevant to measurements
of the vector boson fusion (VBF) production channel \cite{DelDuca:2001fn,Klamke:2007cu,Andersen:2010zx}.
In order to fully understand the challenges in the theoretical modelling of QCD in the presence of vetoes,
and to expose deficiencies in different approaches, observables sensitive to additional radiation 
(such as gap fractions and average jet multiplicities) were measured by ATLAS in refs.~\cite{Aad:2011jz,Aad:2014pua}.
These analyses provided evidence that both high energy and DGLAP logarithms are necessary for 
an adequate description of data.

An algorithm to combine the high energy BFKL logarithms of \HEJ with the soft and collinear logarithms of DGLAP evolution 
was developed for the parton shower \ariadne \cite{Lonnblad:1992tz} in ref.~\cite{Andersen:2011zd}. 
The benefit of such a treatment is not only that the logarithmic accuracies of both descriptions are maintained (such
that emissions under both small and large invariant masses are described
correctly), but also that the partonic results of \HEJ are showered and
hadronised, thus obtaining a more realistic description of the various
stages of a hard scattering. This combined approach compared favourably to
data for several observables. 

Missing from this approach however were two important features that resulted in an inability to correctly describe jet profiles. 
Firstly, the method did not allow for the incorporation of the underlying event \cite{Sjostrand:2004ef}, which is required for a 
successful description of jet profiles. Furthermore, even when the effect of the underlying event was taken into account, 
there was a discrepancy in the profiles of high transverse momentum jets. It transpired that this could be 
understood in terms of certain soft gluons being produced in \HEJ that in \ariadne would only have been produced at a late stage in evolution. 
The presence of such soft emissions thereby inhibited the further evolution of the parton shower and 
such events would not contain the correct amount of collinear radiation.
Although the algorithm properly prevented the double-counting of such soft emissions, there was no mechanism in place
to account for the probability that the parton shower might preferentially have inserted collinear emissions at an earlier stage. 

In this paper we therefore present a new
method for combining the effects of soft and collinear logarithms with 
those of the all-order summation of \HEJ based on the advances made
in the merging of parton showers with fixed-order matrix elements. A crucial feature of our approach 
is that the exclusive $n$-parton events generated according to the \HEJ all-order matrix elements 
will be reweighted using properly subtracted collinear Sudakov factors, and moreover the parton shower will be able to 
insert collinear emissions where it is appropriate to do so. 
This has been implemented for the interleaved parton shower of \py \cite{Sjostrand:2014zea}, allowing for
the inclusion of multiple partonic interactions as well as the
subsequent hadronisation of the event.


The outline of the paper is as follows. The all-order calculation of \HEJ is
described in \cref{sec:hej}. This is followed in \cref{sec:pythia} by a brief description of the
relevant parts of \py and the tree-level matrix element merging procedures from which our algorithm is inspired.
\Cref{sec:matching} will present the method for combining the calculations of \py with \HEJ.
In \cref{sec:results} we examine the performance our merged description, firstly by 
demonstrating the capacity of the new approach to describe jet profiles.
Secondly we compare to data for a set of observables that measure additional radiation in inclusive dijet events. 
We note that in this paper we restrict our focus to pure dijet studies, despite the relevance to Higgs phenomenology.
The reasons for this are two-fold. Firstly, the observables of interest have not yet been measured for Higgs plus dijet processes,
and secondly it is preferable to test newly developed tools in cleaner environments where there is no expectation of new physics.
Nevertheless the method we present should be easily applicable to other processes.
Finally we present the conclusions and outlook in \cref{sec:outlook}.


\section{The High Energy Jets Monte Carlo}
\label{sec:hej}

\subsection{The High Energy Jets Formalism}
\label{sec:hejformalism}
The framework of \emph{High Energy Jets} (\HEJ)~\cite{Andersen:2009nu,Andersen:2009he,Andersen:2011hs} provides an
approximation to the perturbative hard scattering matrix
elements for jet production to any order in the strong coupling. The results are
exact in the limit of large invariant mass between all particles. The
formalism is inspired by the high energy factorisation of matrix elements (as
pioneered by
BFKL~\cite{Fadin:1975cb,Kuraev:1976ge,Kuraev:1977fs,Balitsky:1978ic}), and
obtains a power series in $\hat s$ for the square of the scattering matrix
elements. 
Within \HEJ, approximations are only applied to the matrix elements. This is different to the framework
of BFKL, where numerous kinematic approximations are applied in order to cast
the cross section in the form of a two-dimensional integral equation.
The highest power in $\hat s/p_t^2$ from the square of the matrix element gives the leading-logarithmic
contribution (in $\hat s/p_t^2$) to the cross section. Logarithmic
corrections additionally arise from virtual corrections. Recently it was shown that 
some next-to-leading contributions may be reached within \HEJ \cite{Andersen:2017kfc} 
by including so-called unordered emissions, which have the square of the matrix elements suppressed by one
power in $\hat s$ compared to the leading flavour-configuration for the same
rapidity-ordered momenta. However, in the present study we consider only the leading-logarithmic
contributions to the cross section, where only certain Fadin-Kuraev-Lipatov~\cite{Kuraev:1976ge} 
(FKL) partonic configurations contribute.

We will now discuss in more detail the features of \HEJ that are
relevant to the construction of an algorithm for merging with a parton shower. The all-order
perturbative treatment of  $pp\to jj$ scattering in \HEJ starts with an
approximation to the tree-level amplitude for the scattering
process $f_1 f_2\to f_1 g\cdots gf_2$, where the final-state
particles are listed according to their ordering in rapidity, and $f_1, f_2$ can be
quarks, antiquarks or gluons. These are the FKL configurations that
give rise to the leading contribution to the inclusive $n$-jet cross section in the
\emph{Multi-Regge-Kinematic} (MRK) limit (see ref.\@ \cite{Andersen:2017kfc} for a
recent discussion of the power-suppression of other partonic contributions to
the same multi-jet process). The MRK limit can be specified as the limit of 
large rapidity separations between all particles, for fixed transverse momentum scales:
\begin{align}
 \label{eq:MRKlimt}
\forall i:&&  y_1 \gg \dots \gg y_{i-1}\gg y_i\gg \dots \gg y_n;&& {p_\perp}_i \sim p_\perp
\end{align}
It should be noted that the existence of large transverse momentum
hierarchies is not compatible with the MRK limit, which
will be of importance later.

The
$2\to n$ scattering amplitude is approximated at lowest order by the following
expression~\cite{Andersen:2011hs}:
\begin{align}
  \label{eq:multijetVs}
  \begin{split}
    \left|\overline{\mathcal{M}}^t_{f_1 f_2\to f_1g\ldots gf_2}\right|^2\ =\ &\frac 1 {4\
      (\Nc^2-1)}\ \left\|S_{f_1f_2\to f_1f_2}\right\|^2\\
    &\cdot\ \left(g^2\ K_{f_1}\ \frac 1 {t_1}\right) \cdot\ \left(g^2\ K_{f_2}\ \frac 1
      {t_{n-1}}\right)\\
    & \cdot \prod_{i=1}^{n-2} \left( \frac{-g^2 C_A}{t_it_{i+1}}\
      V^\mu(q_i,q_{i+1})V_\mu(q_i,q_{i+1}) \right),
  \end{split}
\end{align}
where $\left\|S_{f_1 f_2\to f_1 f_2}\right\|^2$ denotes the square of a pure
current-current scattering, $K_{f_1}, K_{f_2}$ are flavour-dependent
colour-factors (which can depend also on the momentum of the particles of
each flavour $f_1, f_2$, see ref.\@ \cite{Andersen:2011hs} for more
details); $q_i$ are the momenta of the colour-octets exchanged in the t-channel, and $t_i=q_i^2$.  The leading-logarithmic contribution to jet production beyond the
first two jets is given by
gluon emission from the underlying $2\to2$ process $f_1 f_2\to f_1 f_2$, and the effective vertex for gluon emissions takes the form \cite{Andersen:2009nu}:
\begin{align}
  \label{eq:GenEmissionV}
  \begin{split}
  V^\rho(q_i,q_{i+1})=&-(q_i+q_{i+1})^\rho \\
  &+ \frac{p_A^\rho}{2} \left( \frac{q_i^2}{p_{i+1}\cdot p_A} +
  \frac{p_{i+1}\cdot p_B}{p_A\cdot p_B} + \frac{p_{i+1}\cdot p_n}{p_A\cdot p_n}\right) +
p_A \leftrightarrow p_1 \\ 
  &- \frac{p_B^\rho}{2} \left( \frac{q_{i+1}^2}{p_{i+1} \cdot p_B} + \frac{p_{i+1}\cdot
      p_A}{p_B\cdot p_A} + \frac{p_{i+1}\cdot p_1}{p_B\cdot p_1} \right) - p_B
  \leftrightarrow p_n.
  \end{split}
\end{align}
This form of the effective vertex is fully gauge invariant; the Ward
Identity, $p_j\cdot V=0$ ($j=2,...,n-1$) can easily be checked, and is
valid for any values for the outgoing momenta $p_j$ (that is, not just in the MRK limit).

The virtual corrections to the amplitude for each multiplicity are
approximated in $D=4+2\varepsilon$ dimensions with the \emph{Lipatov
 ansatz} \cite{Balitsky:1978ic} for the $t$-channel gluon propagators (see
ref.\@ \cite{Andersen:2009nu} for more details). This is obtained by the simple
replacement
\begin{align}
  \label{eq:LipatovAnsatz} \frac 1 {t_i}\ \to\ \frac 1 {t_i}\ \exp\left[\hat
\alpha (q_i)(y_{i-1}-y_i)\right]
\end{align}
in \cref{eq:multijetVs}, where $y_i$ are the rapidities of the outgoing partons and
\begin{align} 
  \hat{\alpha}(q_i)&=-\gs^2\ \Ca\
  \frac{\Gamma(1-\varepsilon)}{(4\pi)^{2+\varepsilon}}\frac 2
  \varepsilon\left({\bf q}_i^2/\mu^2\right)^\varepsilon\label{eq:ahatdimreg},
\end{align} 
is the Regge trajectory which is regulated in $D=4+2\varepsilon$ dimensions, in which
$\mathbf{q}_i^2$ is the Euclidean square of the transverse components of
$q_i$. The cancellation of the poles in $\varepsilon$ between the real and
virtual corrections is organised using a subtraction term,
such that the regulated matrix elements used in the  all-order summation of \HEJ
are given by \cite{Andersen:2011hs}:
\begin{align}
  \label{eq:MHEJ}
  \begin{split}
    \overline{\left|\mathcal{M}_{\rm HEJ}^{\mathrm{reg}, f_1 f_2\to f_1 g
          \cdots g f_2}(\{ p_i\})\right|}^2 = \ &\frac 1 {4\
       (\Nc^2-1)}\ \left\|S_{f_1 f_2\to f_1 f_2}\right\|^2\\
     &\cdot\ \left(g^2\ K_{f_1}\ \frac 1 {t_1}\right) \cdot\ \left(g^2\ K_{f_2}\ \frac 1
       {t_{n-1}}\right)\\
     & \cdot \prod_{i=1}^{n-2} \left( {g^2 C_A}\
       \left(\frac {-1}{t_it_{i+1}} V^\mu(q_i,q_{i+1})V_\mu(q_i,q_{i+1}) \right. \right. \\
     &\left. \left. \phantom{\cdot \prod_{i=1}^{n-2} g^2 C_A \frac {-1}{t_it_{i+1}}}-\frac{4}{\mathbf{p}_i^2} \ \Theta\left(\mathbf{p}_i^2<\lambda^ 2\right)\right)\right)\\
     & \cdot \prod_{j=1}^{n-1} \exp\left[\omega^0(q_j,\lambda)(y_{j-1}-y_j)\right],
  \end{split}
\end{align}
where
\begin{equation}
\omega^0(q_j,\lambda)=\ -\frac{\alpha_s(\mu^2_R) \Ca}{\pi} \log\frac{{\bf q}_j^2}{\lambda^2}.
\end{equation}
and $\lambda$ is a regularisation parameter describing the extent of the
subtraction terms in the real emissions phase space.

Here $\alpha_s$ is evaluated using a renormalisation scale $\mu_R$, which
typically is chosen to reflect the momenta of the final-state
partons. Possible choices include half the scalar sum of transverse momenta
($H_T/2$) and the maximum jet transverse momentum (${p_T}_\mathrm{max}$).
Since the matrix elements have been regulated, this allows for a finite numerical approximation to the
all-order scattering amplitude to be constructed, and for this to be integrated
over all of phase space using a Monte Carlo approach (allowing for the application of arbitrary phase space cuts).
Just as for perturbative fixed-order calculations, the parton momenta in
Eq.~(\ref{eq:MHEJ}) are interpreted as arising from identifiable partons. An
NLO calculation of the production of dijets would deliver the \emph{exclusive} dijet
cross-section to order $\as^3$ and the \emph{inclusive} trijet cross-section
at the same order in $\alpha_s$. The perturbative result in Eq.~(\ref{eq:MHEJ}) contains real
and virtual corrections to any order, and the momenta and multiplicities
should all be considered exclusive (to the logarithmic accuracy of \HEJ).

Indeed, the all-order dijet cross section is  
simply calculated by explicitly summing the exclusive 
$n$-parton cross sections (calculated by numerically integrating the 
matrix elements squared from \cref{eq:MHEJ} over all of phase space)
over all numbers of gluon emissions from 
the initial scattering $f_1 f_2 \to f_1 f_2$.
In addition, matching to tree-level matrix
elements is performed by reweighting each exclusive $m$-jet event
with the factor:
\begin{align}
  w_{m-\mathrm{jet}}\equiv\frac{\overline{\left|\mathcal{M}^{f_1f_2\to f_1g\cdots
          gf_2}\left(\left\{p^\mathrm{new}_{\mathcal{J}_l}(\{p_i\})\right\}\right)\right|}^2}{\overline{\left|\mathcal{M}_{\mathrm{HEJ}}^{t,f_1f_2\to
          f_1g\cdots
          gf_2}\left(\left\{p^\mathrm{new}_{\mathcal{J}_l}(\{p_i\})\right\}\right)\right|}^2}.
\end{align}
This is just the ratio of the square of
the full tree-level matrix element (evaluated using \madgraph \cite{Alwall:2014hca})
to the approximation of this in \cref{eq:multijetVs}, both evaluated on
a set of shuffled momenta $p^\mathrm{new}_{\mathcal{J}_l}(\{p_i\})$ derived from the hard jets only. 
This procedure is summarised in the following formula:
\begin{align}
  \begin{split}
    \label{eq:resumdijetFKLmatched}
    \sigma_{2j}^\mathrm{resum, match}=&\sum_{f_1, f_2}\ \sum_{n=2}^\infty\
    \prod_{i=1}^n\left(\int_{p_{i\perp}=0}^{p_{i\perp}=\infty}
      \frac{\mathrm{d}^2\mathbf{p}_{i\perp}}{(2\pi)^3}\ 
      \int_{y_{i-1}}^{y_\mathrm{max}} \frac{\mathrm{d} y_i}{2}
    \right)\
    \frac{\overline{|\mathcal{M}_{\mathrm{HEJ}}^{\mathrm{reg},f_1 f_2\to f_1 g\cdots gf_2}(\{ p_i\})|}^2}{\hat s^2} \\
    &\times\ \sum_m \mathcal{O}_{mj}^e(\{p_i\})\ w_{m-\mathrm{jet}}\\
    &\times\ \ x_a f_{A,f_1}(x_a, Q^2_a)\ x_2 f_{B,f_2}(x_b, Q^2_b)\ (2\pi)^4\ \delta^2\!\!\left(\sum_{i=1}^n
      \mathbf{p}_{i\perp}\right )\ \mathcal{O}_{2j}(\{p_i\}),
  \end{split}
\end{align}
where $n$ is the partonic multiplicity of the final state, 
and the operator $\mathcal{O}^e_{mj}$ returns one if there are exactly $m$ jets, and zero otherwise.
We also define the inclusive dijet operator $\mathcal{O}_{2j} = \sum^\infty_{n=2} \mathcal{O}^e_{mj}$,
and require that the extremal partons from \HEJ are
members of the extremal jets, in order to ensure that the partonic
configuration matches the situation for which the \HEJ framework was
developed. 
The last line in \cref{eq:resumdijetFKLmatched} corresponds to the inclusion of
parton density functions (PDFs) and the momentum-conserving delta-functional. 
Finally we note that while the sum in the first line of \cref{eq:resumdijetFKLmatched}
is over all numbers of final-state partons, $2 \leq n < \infty$, in practice
the sum needs to include only a finite number of terms: for
finite rapidities and collider energies, the contribution beyond a certain
number of gluons is perturbatively suppressed. The upper bound
$N$ is chosen such that the results are insensitive to further
emissions. This check is performed by simply keeping track of the
contribution from each term in the series, and $N=22$ is sufficient for this study.
Nevertheless, for completeness this choice will enter into the merging algorithm described in \cref{sec:matching}.

The matching of \HEJ to tree-level accuracy is currently
performed up to four jets. The limit on the multiplicity is determined by the time taken
to evaluate the full expressions.  
In addition, the partonic configurations not conforming to the ordering described above
are included in \HEJ by simply adding the contributions order by order (again
using \madgraph \cite{Alwall:2014hca}), but no all-order summation is
performed for these \emph{non-FKL} configurations. 
In the current study, we
will focus on the FKL configurations, since 
this is where special attention is needed in order to avoid double-counting.
This is unlike the challenge addressed by typical fixed-order merging algorithms, 
because the description in \HEJ goes
beyond approximating leading-order matrix elements.
As previously discussed, application of the \emph{Lipatov ansatz} through \cref{eq:LipatovAnsatz} is used to 
sum to all-orders the leading-logarithmic virtual corrections to the $t$-channel poles. 
Although the approaches of \HEJ and \py are complementary and 
calculate different all-order contributions to the perturbative series, 
they cover overlapping regions of phase space, and the
combination of \HEJ and \py therefore requires a new merging algorithm.

We conclude this overview on \HEJ by reiterating that 
a parton shower framework such as \py\cite{Sjostrand:2014zea} is
necessary in order to evolve the partonic state of \HEJ to the state of
hadronisation, primarily by populating the partonic state with further soft
and collinear radiation. In order to obtain the logarithmic accuracy of the shower, it
should also populate (with the appropriate probability) any region between
disparate transverse scales,
which might be generated by \HEJ. Since the shower, as well as the subsequent string
hadronisation, relies on well-defined colour connections between
partons, we now briefly discuss the colour connections arising in \HEJ.

\subsection{The Colour Connections of High Energy Jets}
\label{sec:colo-conn-high}
\begin{figure}
  \hspace{2cm}
  \begin{minipage}[b]{0.5cm}
    \begin{flushright}
      $a$

      \vspace{3.3cm}
      $b$

      \vspace{0.3cm}
    \end{flushright}
  \end{minipage}
  \includegraphics[height=4.5cm]{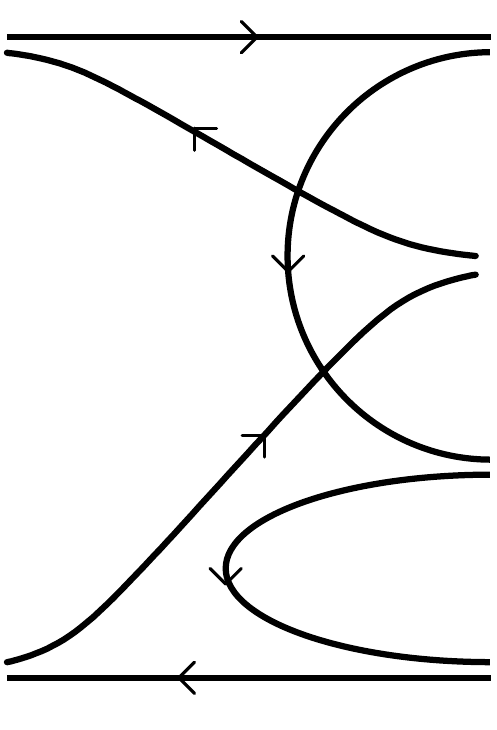}
  \hspace{2.5cm}
  \includegraphics[height=4.5cm]{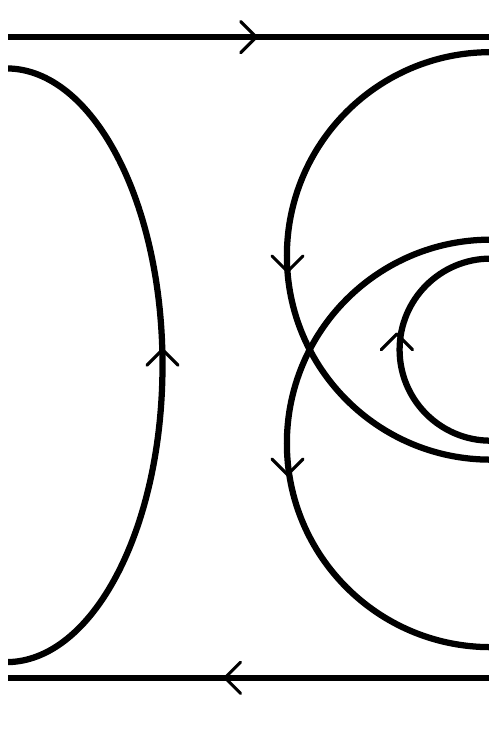}
  \hspace{-6.cm} 
  \begin{minipage}[b]{4cm}
    $1$ \hspace{2cm} $a$

    \vspace{0.8cm}
    $2$

    \vspace{0.7cm}
    $3$

    \vspace{0.7cm}
    $4$ \hspace{2cm} $b$

    \vspace{0.3cm}
  \end{minipage}
  \hspace{1.6cm}
  \begin{minipage}[b]{0.5cm}
    $1$

    \vspace{0.7cm}
    $2$

    \vspace{0.7cm}
    $3$

    \vspace{0.8cm}
    $4$

    \vspace{0.3cm}
  \end{minipage}
  \caption{Examples of a colour flow (left) which contributes in the
    limit of wide angle, hard radiation, and (right) a configuration which
    is suppressed in the same limit. In these diagrams, the final-state
    gluons (on the right of each picture) are ordered according to their
    rapidity.}
  \label{fig:Colours}
\end{figure}

The colour-ordered Parke-Taylor amplitudes~\cite{Parke:1986gb} for tree-level
$gg\to g\cdots g$ scattering allow for a very neat
analysis~\cite{DelDuca:1993pp,DelDuca:1995zy} of the dominant colour
configurations in the limit of widely-separated, hard gluons. The
conclusion, as presented in references \cite{DelDuca:1993pp,DelDuca:1995zy}, is
that the leading contribution in the MRK limit is provided by the colour configurations which can be untwisted
into two \textit{non-crossing} ladders that connect the rapidity-ordered
gluons. \Cref{fig:Colours} (left) contains an example of a configuration
contributing in the MRK limit, and one (right) which is suppressed. The
numbering of the final-state partons is assigned according to their rapidity; as drawn their vertical ordering also 
coincides with their ordering in rapidity.

The colour connections in \cref{fig:Colours} (left) can be summarised as
$a134b2a$. It is possible to arrange the final-state partons such that no colour lines cross 
without modifying the vertical order of the final-state particles, namely by moving particle 2
to the left side of the same plot. Since the vertical ordering is unchanged,
the rungs of the resulting
un-crossed ladders are also ordered in rapidity.
Such manoeuvres are always possible when the order of the particles 
in the colour connection string between the two
initial-state gluons $a \dots b$ and $b \dots a$  
also reflects their order in rapidity, as in the case of $\{134\}$ and $\{2\}$
in $a134b2a$. 

The colour connections in \cref{fig:Colours} (right) can be
summarised as $a1324ba$; in this case the string $\{1324\}$ between
 $a$ and $b$ is \textit{not} ordered in rapidity. The only manoeuvre
 which will untangle the colour connections requires flipping the vertical arrangement of
particles $2$ and $3$ such that their vertical ordering is no longer equivalent to their ordering in rapidity. 
This configuration is therefore suppressed in the
MRK limit, because the two un-crossed ladders are not rapidity-ordered.

Furthermore, the study of references \cite{DelDuca:1993pp,DelDuca:1995zy} shows
that all the leading configurations each have the same limit in the
MRK limit, resulting in a colour
factor $\Ca$ for every final-state gluon. The limit agrees with that
predicted by the amplitudes of Fadin-Kuraev-Lipatov (FKL)~\cite{Kuraev:1976ge}.
When we pass an event from \HEJ to \py, we choose a
colour configuration at random from the set of colour connections which are
leading in the MRK limit, and pass the event using an interface conforming
to the \emph{Les Houches accord}~\cite{Boos:2001cv}. This method is identical
to that applied in ref.\@ \cite{Andersen:2011zd}.

\section{\protect \py and CKKW-L}
\label{sec:pythia}

There are several reasons for using \py to handle the collinear
resummation rather than \ariadne as was done in
ref.\@  \cite{Andersen:2011zd}. First of all, the handling of initial-state
radiation in \ariadne is somewhat peculiar \cite{Andersson:1988gp} and
does not quite fit into a conventional resummation
scenario. Furthermore, \py has a much more advanced infrastructure for
handling matching and merging. Finally, \py has a very advanced model
for multiple partonic scattering (based on ref.\@  \cite{Sjostrand:1987su}) which is
needed to have a realistic description of the underlying event.

\subsection{The Interleaved Shower in \py}
\label{sec:interl-show-py}

\py implements a transverse-momentum-ordered shower
\cite{Sjostrand:2004ef}, which includes not only initial- 
and
final-state emissions, but also interleaves these with multiple
partonic scatterings. The general philosophy is that emissions (or
sub-scatterings) with high transverse momentum should always be
performed before those with lower transverse momentum.

As in all parton shower algorithms the ordering is used to ensure that
the probability for any emission remains finite, and that the whole
shower process is unitary. Even though the splitting functions for an
emission diverge for small transverse momenta according to
$P(k_\perp^2,z)\propto1/k_\perp^2$, at each step of the shower the
basic splitting probabilities are amended by the probability that no
splittings with larger transverse momenta had happened before.
The probability that the \emph{hardest} emission occurs at the scale 
$k_\perp^2$ with an energy splitting $z$ is given by:

\begin{align}
  \label{eq:hardest}
  \frac{d\mathcal{P}_{\mbox{\scriptsize hardest}}}{dk_\perp^2dz}=&
  P(k_\perp^2,z)
  \exp\left\{-\int_{k_\perp^2}^{{k^2_\perp}_{\max}}dk_\perp^{'2}
    \int dz' P(k_\perp^{'2}, z')\right\} \nonumber \\
    \equiv &
  P(k_\perp^2,z)\Delta({k^2_\perp}_{\max},k_\perp^2). 
\end{align}
Here the Sudakov factor  $\Delta({k^2_\perp}_{\max},k_\perp^2)$ corresponds to the no-emission probability,
ensuring that there
were no other emissions between the maximum scale ${k^2_\perp}_{\max}$ and $k_\perp^2$. 
A lower cutoff,
${k^2_\perp}_\mathrm{cut}$, is still needed but can be taken very small and
still result in probabilities below unity.
Formally, \cref{eq:hardest} resums the leading double-logarithms of ${k^2_\perp}_{\max}/k_\perp^2$
in the soft-collinear limit in the leading colour (large $N_c$) approximation. 
It should be noted however that many formally subleading contributions, such as momentum conservation,
which in practice give rise to large effects are also included.

The no-emission probabilities are fairly easily implemented using the
Sudakov veto algorithm \cite{Sjostrand:2006za}, and
has simple factorisation properties if several different types of
emissions are possible, due to the exponential form. The ordering variable, $k_\perp$, used 
in the evolution is not necessarily the actual transverse momentum of an emission in any
Lorentz frame, and it is defined slightly differently depending on the class of emission
in the interleaved shower. For final-state radiation -- \textit{FSR} -- (or time-like splittings) it is
defined as ${k^2_\perp}_\mathrm{FSR}=z(1-z)Q^2$, where $Q^2$
is the invariant mass of the two final-state partons. For initial-state radiation -- \textit{ISR} --
(or space-like splittings) we instead have ${k^2_\perp}_\mathrm{ISR}=(1-z)Q^2$, where now $Q^2$ is the virtuality of
the incoming parton entering the hard sub-system after the emission. Finally for
multi-parton interactions (\textit{MPI}), ${k^2_\perp}_\mathrm{MPI}$ is simply
defined as the transverse momentum in the lab system for the $2\to2$
scattering.

\subsection{Merging \`a la CKKW(-L)}
\label{sec:merging-a-la}

The partonic states generated by a parton shower are exclusive;
in other words, the probability to produce an $n$-parton state in the parton shower is approximately given by the exclusive
cross section for exactly $n$ partons.
This is in contrast to
$n$-parton states generated by a matrix element generator,
where the state is exactly given by the inclusive cross
section for having \emph{at least} $n$ partons. The main principle
of algorithms that merge matrix elements with parton showers is
therefore to take several inclusive samples with different numbers of
partons from a matrix element generator and reweight them with
no-emission probabilities to make them exclusive. 
This allows the
samples to be safely added and subsequently showered 
without any double-counting.

The general idea in this paper is to use \HEJ as a matrix element
generator and add emissions from \py in a consistent way. In doing so
we will use ideas from the CKKW-L merging algorithm
\cite{Lonnblad:2001iq,Lavesson:2005xu,Lonnblad:2011xx}, but with some
important modifications which will be described in \cref{sec:matching}.
Here we shall review the pertinent features of the CKKW-L method. 

Similarly to
merging algorithms such as CKKW \cite{Catani:2001cc} and
MLM \cite{Mangano:2006rw}, the CKKW-L method takes matrix-element-generated states
and tries to reconstruct a sequence of emission scales from which the
no-emission probabilities are calculated. While some merging procedures use
jet clustering algorithms to do this, CKKW-L looks at the partonic states and
tries to answer the question \textit{``How would my parton shower have
  generated this state?''}, and then reconstructs the full kinematics of
the corresponding sequence of emissions in the parton shower. Often
there is more than one sequence of emissions possible, in which case
one sequence is chosen at random with relative weights given by 
the product of the values of the corresponding splitting
functions. The sequence chosen is referred to as the parton shower history
and will comprise of a complete set of intermediate states,
$\{S_0,\ldots,S_n\}$ (where $S_0$ is the lowest multiplicity state and
$S_i$ has $i$ additional partons) and a series of $n$ parton shower
emissions. Each emission $i$ is characterised by an ordering scale
${k^2_\perp}_i$, a splitting fraction $z_i$, and an azimuthal angle,
$\phi_i$. This procedure differs from the standard
  CKKW algorithm, where the intermediate states are not needed and
  instead only the emission scales are calculated by the
  $k_\perp$ jet-clustering algorithm. Formally this difference only affects
  sub-leading logarithms.

The no-emission probabilities are then
calculated by generating trial emissions from each intermediate state in turn, starting at
$S_0$. The emission generated from $S_i$ will have a maximum
scale given by ${k^2_\perp}_i$. The probability that this emission has a
scale above ${k^2_\perp}_{i+1}$ is exactly the no-emission probability
$\Delta_i({k^2_\perp}_i,{k^2_\perp}_{i+1})$. Giving the
matrix-element-generated state a weight zero if a trial
emission from a given state $S_i$ has a scale above ${k^2_\perp}_{i+1}$
is therefore equivalent to reweighting the cross
section by the no-emission probability:
\begin{equation}
  \label{eq:noem}
  \prod_{i=0}^n\Delta_i({k^2_\perp}_i,{k^2_\perp}_{i+1}).
\end{equation}
Here ${k^2_\perp}_0 $ is the maximum possible
scale and corresponds to the scale of the Born level process; ${k^2_\perp}_{n+1}\equiv {k^2_\perp}_M$ is the merging
scale which is given by the cut used in the matrix element
generator, and is used to isolate the region of soft and collinear divergences
where the parton shower is assumed to give a better description.

We can now freely add more partons below the merging scale with our
parton shower.  For the case that $n=N$ is the maximum multiplicity of
the matrix element samples to be merged, trial emissions from $S_N$
are not performed and the last factor
$\Delta_N({k^2_\perp}_\mathrm{N},{k^2_\perp}_{N+1})$ is omitted. (This
is because there is no possibility of double-counting with states of
higher multiplicity, $n>N$.)  Consequently the shower is instead
started from ${k^2_\perp}_N$.

In addition to the reweighting of the cross section by the no-emission probability,
there is also a reweighting of the value for \as\ used in the matrix element,
typically evaluated at some fixed renormalisation scale $\mu_R$ characteristic of the Born level process. For a parton
shower resummation, however, it can be shown that true collinear
logarithms are better reproduced if \as\ is evaluated at the
scale of the individual shower splittings. The states
are therefore reweighted by the factor: 
\begin{equation}
\frac{1}{\as^n(\mu_R^2)}\prod_{i=1}^n \as({k^2_\perp}_i).
\label{eq:alpharew}
\end{equation}
Additionally there is a reweighting with PDFs,
related to the fact that the no-emission probabilities contains PDF
ratios, as explained in more detail in \cite{Lavesson:2005xu}.
The end result is that the merged event sample will be constructed by
exclusive partonic states where the $N$ hardest emissions above the
merging scale are given by the full tree-level matrix element, and all
softer emissions are given by the shower.

So far we have only considered initial- and final-state showers, but
to get a realistic description we also need to consider the underlying
event. This cannot be described by a tree-level matrix element, but it
may be accounted for in an interleaved parton shower using MPI.
This means that we also want to incorporate the MPI ``emissions''
from the shower in the merged event sample. 
The underlying event may actually contain hard
jets, and it is impossible to separate these from the jets given by
the matrix element generator. Therefore we cannot blindly include MPI
emissions in the CKKW-L no-emission probabilities above, but the
procedure is modified \cite{Lonnblad:2011xx} as follows.

As before we reconstruct a parton shower history for every matrix
element state. We make trial emissions \textit{including MPI} 
from each intermediate state $S_i$ for $i<n$,
giving the event a weight zero if the emission scale is above ${k^2_\perp}_{i+1}$.
The last state $S_n$ is treated separately. 
When an emission is generated above ${k^2_\perp}_M$, if it corresponds to either
ISR or FSR we still give the event a weight zero; if
however an MPI is generated we will accept the generated state and
continue the shower below the emission scale rather than below the merging
scale.
The end result is thus changed such that the merged event sample will now
consist of exclusive partonic states where the $N$ hardest emissions
above the merging scale \emph{that are not from an MPI} are given by
the full tree-level matrix element, and all softer emissions are given by
the shower. 

  We note that merging procedures such as CKKW-L are not necessarily
  unitary, in that the inclusive lowest multiplicity Born level cross
  section is not preserved, as is the case in parton showers. This is
  because of the mismatch between the ratio of full matrix element
  describing the addition of a parton and the splitting function used
  in the no-emission probabilities. This is in contrast to
  \emph{matching} procedures (see \eg~\cite{Bengtsson:1987rw}) where
  it is the matrix element ratios that are exponentiated in the
  no-emission probabilities.

%
%


\def\sigmab{\ensuremath{\sigma_2^{\star}}}

\section{Modified Merging for \HEJ}
\label{sec:matching}

In merging matrix elements with parton showers there are two primary challenges encountered, which we recapitulate here so as to compare the corresponding 
challenges in merging \HEJ with a parton shower. The first challenge is to ensure there is no double-counting between the fixed order matrix elements and the parton shower.
In fixed order merging algorithms, this is achieved through the merging scale, which provides a clear partition of phase space. Above the merging scale, the multiplicity
of hard jets should not be increased by the parton shower, and the distribution of hard jets should be determined by the fixed order matrix elements.
Below the merging scale, soft and collinear radiation from the parton shower is added, smearing the energy of the original hard partons, but leaving the 
original jets' energies largely unchanged.

We want the merging of \HEJ and \pyt to obtain the logarithmic accuracy of
both. Therefore, the parton shower should not change the jet multiplicity relative to 
\HEJ in the MRK limit (namely, at large rapidities with no transverse momentum hierarchies).
The parton shower should however be able to add collinear radiation inside the jet cone. 
One could envisage using a phase space slicing mechanism such that
regions populated by \HEJ and parton shower are not allowed to
overlap.
However, in combining \HEJ with a parton shower 
we are aiming to correctly model the amount of radiation (for example, the multiplicity of jets)
in regions of phase space sensitive to both high energy and soft-collinear logarithms, which is hard to achieve with a strict partition.
Instead we will allow both
formalisms to populate their respective overlapping phase spaces and define a subtraction
term for the splitting functions and corresponding no-emission probabilities in the shower. 
Double-counting is then avoided by reducing
the probability of producing a certain emission in the shower by the probability that
\HEJ had already performed that emission. 

The second challenge in fixed order
  merging is to avoid double-counting between the inclusive event
  samples that are combined, which is resolved by making those event
  samples exclusive through reweighting with Sudakov factors.  The
  picture for merging \HEJ with a parton shower is slightly different,
  because a given $n$-parton event generated by \HEJ is already
  exclusive. (This is ensured by the inclusion of virtual corrections
  at the level of the matrix elements to all orders in
  \cref{eq:MHEJ}.)  It would therefore be inappropriate to na\"ively
  reweight events with the Sudakov factors in \cref{eq:noem} whose kernels correspond to
  the full Altarelli-Parisi splitting functions.
Instead we have devised a procedure where
  collinear emissions from the shower are added to states produced by
  \HEJ in a way such that the corresponding collinear Sudakov form factors only
  change the relative weight of different \HEJ multiplicities, and retain the
  inclusive cross section. It does so by inserting emissions also in
  early stages of the reconstructed parton shower history to avoid
  \textit{under}-counting of collinear emissions due to phase space
  limitations set by the full generated \HEJ state, which was the main
  drawback of the previous approach in \cite{Andersen:2011zd}.

In \cref{sec:ckkw-l-hej} we will outline the merging procedure without specifying the particulars of how the division of phase space
between \HEJ and parton shower is achieved. We simply assume that there exists a consistent way to classify a given emission as being 
belonging to the either the \HEJ or parton shower regimes. 
We use the jet cone radius as an example of a cut 
to highlight some features of our algorithm.
Interpreting this statement in the language of the parton shower
implies that it is possible to define both \HEJ and collinear (or subtracted) Sudakov factors. 
We will then develop these ideas, in particular the
definition of and procedure to calculate these subtracted Sudakov factors in \cref{sec:subtracted}.
Finally, the algorithm will be disclosed in full in \cref{sec:ModifiedMergingAlgo}.

\subsection{CKKW-L and \HEJ}
\label{sec:ckkw-l-hej}

A prescription for dressing \HEJ events with collinear radiation may be 
obtained in a manner analogous to how MPI were added to samples of tree-level
events in CKKW-L. To understand how this works, we first note that the MPI algorithm
could have been reformulated such that one first does a normal reweighting
with the no-emission probabilities \textit{excluding MPI} in the trial
emissions, and then go through the reconstructed states
a second time making trial emissions \textit{using only MPI}. In this second
round, starting from $S_0$, as soon as an MPI trial emission from
$S_i$ with a scale $k_\perp > {k_\perp}_{i+1}$ is found one simply
replaces the original $S_n$ state with the $S_i$ state plus the
additional generated MPI emission. The shower subsequently evolves from 
the MPI emission scale $k_\perp$.

In the analogous procedure for \HEJ, since the states are already
exclusive we completely skip the first round of reweighting, and
proceed directly to the adding of collinear splittings and MPI.  As
before this is done by first constructing the parton shower history,
however the reconstructed states should only correspond to
configurations which \HEJ could have generated. We will define such
`\HEJ states' more precisely in the next section.  This is followed by the generation
of trial emissions from each reconstructed state \textit{which \HEJ
  could not have done}, namely the collinear emissions and MPI. 

If a trial emission from state $S_i$ is generated that has 
a scale $k_\perp > {k_\perp}_{i+1}$, the original
$S_n$ state is replaced by the reconstructed state $S_i$ with the
additional trial emission. If the original event is replaced, the
shower is allowed to evolve freely from the scale $k_\perp$.  In such
a prescription, the $N$ hardest emissions that are neither collinear
nor correspond to an MPI are generated by \HEJ; everything else is
generated by the shower.  We also skip the reweighting of \as\ in
\cref{eq:alpharew}, since as discussed in \cref{sec:hej} the scale
used in \HEJ has been chosen to be characteristic of the event
topology.  We will however still use $\as(k_\perp^2)$ in the addition
of collinear emissions from the shower.

\def\kti#1{\ensuremath{{k^2_\perp}_{#1}}}
To illustrate how the
  algorithm works quantitatively we will in the following assume that
  there is a clean separation in phase space between the \HEJ states
  and the region where we want to dress the jets with collinear
  emissions from \pyt. For instance, we could imagine a simple phase space
  cut, where \HEJ states are required to have a $\Delta R$ between any two partons
  larger than some value, and the \pyt splitting functions are set to
  zero if they result in such states.

We start by reformulating the $n$-parton state produced by \HEJ within the
shower-formalism, written as a basic two-parton inclusive cross section
multiplied by a series of `\HEJ splittings' with decreasing values of
the scale reconstructed by the merging algorithm.
%
%
The fact that
the \HEJ states are exclusive means we can write the cross section for an
$n$-parton state produced by \HEJ (that is, prior to merging) as:
\begin{equation}
  \label{eq:HEJ-history}
  d\sigma_n^H=
  d\sigmab \left(\prod_{i=3}^nP_{i\!-\!1}^H(\kti{i})\Delta_{i\!-\!1}^H(\kti{i\!-\!1},\kti{i})
    d\kti{i}\Theta(\kti{i\!-\!1}-\kti{i})\right)\times\Delta_{n}^H(\kti{n},\kti{M}).
\end{equation}
Here $P^H_i(k_\perp^2)$ is the splitting function for
emitting a parton at the scale $k_\perp^2$ from the state $i$
\textit{according to \HEJ}, integrated over the energy fraction $z$;
$\Delta_i^H(\kti{i},k_\perp^2)$ is the probability that there were no
`\HEJ-like' emissions from the state $i$ between the scales $\kti{i}$
and $k_\perp^2$.  Finally, $d\sigmab$ is the inclusive differential
cross section for the initial two-parton state. 

The shower-merging will have to construct all shower-histories, which could
have produced a given $n$-parton state from \HEJ. This ends up being the
time-consuming step for the high-multiplicity states produced by \HEJ. These
states can be of much higher multiplicity than the current limit experienced
with fixed-order matchings, where the shower-histories are also
reconstructed. In order to reduce the complexity of the shower history
reconstruction, we trim the parton-content of the high-multiplicity
states from \HEJ before they are passed to the shower. This is done removing any parton with a
transverse momentum smaller than a scale $k_{\perp M}$ from the event record
(and reshuffling the remaining momenta to absorb the transverse momentum thus
removed).  The effect of
introducing the trimming though is that the event record contains no partons
with transverse momenta less than $k_{\perp M}$. After the trimming, this
phase space is therefore left completely for the shower to populate, and the
trimming scale is thus the final scale for the last Sudakov factor in
Eq.~(\ref{eq:HEJ-history}).

The inclusion of trimming can speed up the merging significantly;
however, it should be emphasised that \textit{formally} we should consider the 
limit where $k_{\perp M}\to 0$. Nevertheless, as the weight of the event is kept unchanged,
as long as $k_{\perp M}$ is smaller than the scale of the jet threshold, any observable
based on jet momenta is only weakly dependent on this trimming if at
all. We will later (in figure \ref{fig:avejetspure})
investigate directly the numerical impact of the transverse scale used in the
trimming of the event record in passing events from \HEJ to \pyt, which
indeed is found to be insignificant even on the observables which are very
sensitive to the jet multiplicities of the events.

Before continuing,
some comments may be needed to clarify \cref{eq:HEJ-history}:
\begin{itemize}\itemsep 0mm
\item At this stage we do not need to know anything about $P^H$. The
  fact that we have a sequence of emissions means that we can describe
  it as a product of splitting functions accompanied by corresponding
  no-emission probabilities which are of the form given in
  \cref{eq:hardest}, even if the states were not produced that way by
  \HEJ.
\item In rare cases it is not possible to reconstruct an ordered
  history of shower emission. Such cases are handled by joining two
  (or more) subsequent steps into one, as described
  in~\cite{Lonnblad:2011xx}. Such unordered paths are by definition
  far the parton shower resummation regions and does not affect the
  logarithmic accuracy of the procedure.
\item In other rare cases, it is not possible to find intermediate
  states corresponding to \HEJ states. Again we treat these by joining
  several steps into one, so that the trial emissions always come
  from \HEJ-like states.
\item The total inclusive dijet cross section is given by $\sigmab$ and
  is not the basic tree-level $2\to2$ cross section. This is because
  \HEJ includes non-unitary corrections beyond leading order, and
  these we would like to preserve.
\end{itemize}
We shall now consider the different possibilities which may arise in the merging procedure described above.
In the first case, the original state generated by \HEJ is not replaced by one
generated by the shower. This will occur only if at each reconstructed state in the history,
no trial (non-\HEJ-like) emission is generated above the scale ${k_\perp}_{i+1}$ of the next reconstructed state.
This corresponds to multiplying \cref{eq:HEJ-history} by the following product of no-emission factors:
\begin{equation}
\label{eq:collinear_sudakovs}
\Delta^C_n(\kti{n},\kti{M}) \cdot \prod_{i=3}^n \Delta^C_i(\kti{i\!-\!1},\kti{i}),
\end{equation}
where $\Delta^C_i$ is a suitably modified (collinear) Sudakov
  factor, for example, corresponding to the exponentiation of a \pyt splitting
  function with the $\Delta R$ cut assumed above. 
Defining
$\Delta^M_i=\Delta^H_i\Delta^C_i$, such events will contribute to the cross section according to:
\begin{equation}
  \label{eq:mergems}
  d\tilde{\sigma}_n=
  d\sigmab \left(\prod_{i=3}^nP_{i\!-\!1}^H(\kti{i})\Delta^M_i(\kti{i\!-\!1},\kti{i})
    d\kti{i}\Theta(\kti{i\!-\!1}-\kti{i})\right)\times\Delta^M_n(\kti{n},\kti{M}).
\end{equation}
Furthermore it is clear that we can freely dress these states with full \pyt
splittings below the merging scale.

If instead a trial emission is generated from a reconstructed state
$m\le n$ at the scale $\kti{C}>\kti{M}$ (and above the scale of the
next reconstructed state) the original $n$-parton state from \HEJ will
be replaced by the reconstructed $m$-parton state plus the accepted
trial emission. Calculating the contribution to the cross section from
such states requires summing and integrating over all possible
reconstructed \HEJ emissions below $\kti{C}$,
\begin{align}
  \label{eq:mergeall}
  d\tilde{\sigma}_{m/n}=&d\sigmab \left(\prod_{i=3}^mP^H_{i\!-\!1}(\kti{i})
    \Delta^M_{i\!-\!1}(\kti{i\!-\!1},\kti{i})d\kti{i}\Theta(\kti{i\!-\!1}-\kti{i})\right)
  \times\nonumber\\
  &\times P^C_m(\kti{C})\Delta^M(\kti{m},\kti{C})d\kti{C}\Theta(\kti{m}-\kti{C})\\
  &\times \sum_{j=m+1}^{n} \int_{\kti{M}}^{\kti{C}}\text{all \HEJ emissions}, \nonumber
\end{align}
where $P^C_m$ is the $\Delta R$-truncated \pyt splitting function. For
$n=m$ there is no integral and we just get the probability that there
are no extra emissions. For $n=m+1$, we get the probability that there
is exactly one extra emission, for $n=m+2$ exactly two, etc. The sum
of all these must necessarily sum up to unity, and the final result is
just the two first lines of \cref{eq:mergeall}.

In practice there is an upper cut, $N$, on the parton multiplicity in
\HEJ. Assuming that the corresponding cross section is inclusive over
the last emission, the last integral in the third line of
\cref{eq:mergeall}, becomes
\begin{equation}
  \label{eq:mergelast}
  \int_{\kti{M}}^{\kti{N\!-\!1}}d\kti{N}
  P^H_{N\!-\!1}(\kti{N})\Delta_{N\!-\!1}^H(\kti{N\!-\!1},\kti{N})=
  1-\Delta_{N\!-\!1}^H(\kti{N\!-\!1},\kti{M}),
\end{equation}
where we have used the property of no-emission probabilities that its
derivative is simply itself times the splitting function,
$\frac{d}{dk_\perp^2}\Delta_{i}(\kti{i},k_\perp^2)=P_{i}(k_\perp^2)\Delta_{i}(\kti{i},k_\perp^2)$. When adding this to
the $N-1$ contribution, this will explicitly cancel the last
no-emission factor there, and we can do the last integral in the $N-1$
contribution in the same way, and so on, until we cancel also the last
no-emission factor in the $m=n$ contribution.

Adding full \pyt shower splittings below \kti{C}, we can now write the
exclusive probability that we have exactly $n$ partons above the merging
scale as
\begin{eqnarray}
  \label{eq:mergingtotal}
  d\tilde{\sigma}_{n}&=&
  d\sigmab \left(\prod_{i=3}^nP_{i\!-\!1}^H(\kti{i})\Delta^M_i(\kti{i\!-\!1},\kti{i})
    d\kti{i}\Theta(\kti{i\!-\!1}-\kti{i})\right)\times\Delta^M_n(\kti{n},\kti{M})\nonumber\\
  &+&\sum_{m=3}^{n-1}d\sigmab 
  \left(\prod_{i=3}^mP^H_{i\!-\!1}(\kti{i})
    \Delta^M_{i\!-\!1}(\kti{i\!-\!1},\kti{i})d\kti{i}\Theta(\kti{i\!-\!1}-\kti{i})\right)
  \times\nonumber\\
  & &\qquad\times P^C_m(\kti{C})\Delta^M(\kti{m},\kti{C})d\kti{C}\Theta(\kti{m}-\kti{C})\\
  & &\qquad\times \left(\prod_{i=m+1}^nP^P_{i\!-\!1}(\kti{i})
    \Delta^P_{i\!-\!1}(\kti{i\!-\!1},\kti{i})d\kti{i}\Theta(\kti{i\!-\!1}-\kti{i})\right)
  \times\Delta^P_n(\kti{n},\kti{M}),\nonumber
\end{eqnarray}
where $P^P$ is now the full \pyt splitting function (possibly also
including MPI) and $\Delta^P$ the corresponding no-emission
probability. Comparing this with what \pyt would give on its own,
\begin{equation}
  \label{eq:justpythia}
  d\sigma_n^P=
  d\sigmab \left(\prod_{i=3}^nP_{i\!-\!1}^P(\kti{i})\Delta_{i\!-\!1}^P(\kti{i\!-\!1},\kti{i})
    d\kti{i}\Theta(\kti{i\!-\!1}-\kti{i})\right)\times\Delta_{n}^P(\kti{n},\kti{M}),
\end{equation}
we see that for $n$ partons above the merging scale the $m$ hardest
ones will always be produced by \HEJ, and if there are partons from
\pyt above the merging scale the hardest one of these will always be a
collinear splitting.
We also see that the procedure is unitary,
  in that the inclusive jet cross section is still given by $\sigmab$
  as calculated by \HEJ. All we have done is to add (unitary) parton
  shower emissions and, in some cases where these are harder than the
  \HEJ ones, reclustered the original \HEJ state into a lower
  multiplicity state, and then added the parton shower.

We note that the action of multiplying by \cref{eq:collinear_sudakovs}
was not present in the algorithm presented in \cite{Andersen:2011zd}
for matching \HEJ with \ariadne. That is to say, the probability that
the parton shower might have produced a collinear emission at an
earlier stage in the reconstructed history was not taken into
account. It was the lack of this step which allowed the inclusion of
soft gluons from \HEJ that interfered with the ordering of the parton
shower and prevented a proper parton shower evolution in the full
phase space. Furthermore these collinear emissions which according to
the parton shower should have occurred were not inserted; instead such
emissions could only be included \textit{below} the matching scale.
We emphasise that in this regard the approach we take here is
fundamentally different from how \HEJ was matched with \ariadne.

\subsection{The Subtracted Shower}
\label{sec:subtracted}

%
In the previous subsection it was assumed that we could make a simple
phase space cut between collinear splittings to be described by \pyt
and large angle splittings from \HEJ. However, the MRK-limit in \HEJ
does not take into account large logarithms that arise in case we have
large transverse momentum hierarchies between (possibly widely
separated) jets. Such logarithms are included in the parton shower,
and we would like to include them in our merging.

To accomplish this we go beyond a simple phase space cut and use
\textit{subtracted} splitting functions instead. The idea is that
where resummation is important the splitting functions are large, and
we could naively say that where the \pyt splitting function is less
than the \HEJ one we set it to zero, and vice versa. This would still
correspond to a simple phase space cut, albeit more complicated than
the $\Delta R$ cut assumed above. However, this would be fairly
wasteful as we would throw away many of the jets produced by
\HEJ. Instead we have introduced a procedure where the \HEJ splitting
function is subtracted from the \pyt one.

To do this it is now necessary to obtain an explicit definition of the
splitting functions and no emission probabilities for \HEJ.  Although
no such expressions appear explicitly within the \HEJ formalism, we
note that the Altarelli-Parisi splitting functions may be derived as
the soft and collinear limit of a ratio of matrix elements
\cite{Ellis:1991qj}:
\begin{equation}
  d k_\perp^2 dz  \int d\phi \frac{1}{16\pi^2}
  \frac{|\mathcal{M}^{n+1}|^2}{|\mathcal{M}^n|^2}\sim 
  \frac{d k_\perp^2}{k_\perp^2} dz \frac{ \alpha_s}{2\pi} P_{gg}(z)\;. \label{eq:ME2ratio}
\end{equation}
This is just the normal universal behaviour of
  matrix elements in the soft and collinear limit. The 
  Altarelli-Parisi splitting functions precisely capture the 
  soft and collinear singularities which must be exponentiated to
  calculate the leading DGLAP logarithms in the parton shower no emission probabilities.
  If instead we replace the full matrix elements by the \HEJ
  ones, this will no longer contain any collinear singularities, but only
  the soft singularities. Such a function is precisely what is needed
  to define a subtraction term for the parton shower. 
  Of course, we could take the MRK limit of this and retain
  the same logarithmic accuracy, but by using the full matrix elements we
  retain more of the \HEJ accuracy.

Therefore, as in the approach of \cite{Andersen:2011zd}, we define the \HEJ splitting function as a ratio of \HEJ matrix elements given by \cref{eq:MHEJ} corresponding to 
an event before and after the insertion of an emission as generated by the parton shower. Of course as noted in \cref{sec:hej}, 
these matrix elements are only valid for FKL configurations, but there is no restriction upon the kind of configuration which may be generated by the parton shower. 
We must therefore assert that the following criteria define a `HEJ state':
\begin{enumerate}\itemsep -0.5mm
 \item The most forwards outgoing parton should have the same flavour as the parton incoming along the positive z axis.\label{fklcondition1}
 \item The most backwards outgoing parton should have the same flavour as the parton incoming along the negative z axis. \label{fklcondition2}
 \item All other outgoing partons must be gluons. \label{fklcondition3}
 \item It must be possible to untangle the colour connections into two `ladders' of rapidity-ordered partons.\label{colcondition}
 \item The outgoing partons must cluster into at least two jets. \label{jetcondition1}
 \item Each extremal (most forwards or backwards) parton must be a member of the corresponding extremal jet.\label{jetcondition2}
 \item Each parton must have a transverse momentum above the merging scale ${k_\perp}_M$. \label{mergingscalecondition}
\end{enumerate}
Criteria \ref{fklcondition1}-\ref{fklcondition3} simply define an FKL
configuration; criterion \ref{colcondition} is required since a given
set of colour connections is chosen (as described in
\cref{sec:colo-conn-high}) and this has an impact upon which dipoles
arise in the \pyt parton shower; criteria
\ref{jetcondition1}-\ref{jetcondition2} are kinematic constraints on
\HEJ events.  Finally, although not strictly necessary for the purpose
of efficiency events generated by \HEJ containing soft emissions below
$\kti{M}$ are reclustered (in a manner that does not alter the
rapidities of the resulting jets) and this is therefore reflected by
the requirement given in condition \ref{mergingscalecondition}.  The
reclustering reduces the complexity of constructing all possible
shower histories for the state passed to \pyt. This reduces the CPU
time needed to obtained merged predictions, and explicit tests
indicate that the impact of the reclustering is unnoticeable on
observables based on hard jets, such as those studied in this paper.

Now, for any emission resulting in a configuration not corresponding to a \HEJ state we set
$P^H = 0$ (because there is nothing to subtract for non-FKL states),
and otherwise we can define:
\begin{equation}
P^H  = \frac{1}{2} \frac{1}{16 \pi^2} \frac{|\mathcal{M}^{n+1}_\mathrm{HEJ}|^2}{|\mathcal{M}^n_\mathrm{HEJ}|^2}\;.\label{eq:hej_split}
\end{equation}
The factor of $\frac{1}{2}$ accounts for the fact that the matrix elements are summed over all possible colour connections, but for each parton shower emission we wish to calculate the splitting function for
one of two possible choices (each of which contribute equally in the MRK limit). This expression however only accounts for time-like emissions. 
For a space-like branching $i\rightarrow jk$, where parton $j$ is evolved backwards to parton $i$ with a higher momentum fraction $x_i = (1/z) x_j$, we instead define:
\begin{equation}
\label{eq:hej_split_space}
P^H_\mathrm{spacelike} = \frac{1}{2}  \frac{1}{16 \pi^2}  \frac{|\mathcal{M}^{n+1}_\mathrm{HEJ}|^2}{|\mathcal{M}^n_\mathrm{HEJ}|^2} \, \frac{x_i f_i(x_i,\mu_F^2)}{x_j f_j(x_j,\mu_F^2)}\;.
\end{equation}
where the PDFs $f_{i,j}$ should be evaluated at an appropriately
chosen factorisation scale $\mu_F$.  Our effective Sudakov factor $\Delta^H$
from the previous section would then simply be the exponentiation of this
splitting kernel, however, we will not need to compute this explicitly in the numerical
implementation of the algorithm.
For completeness we shall also write down the \py
splitting functions, evaluated as a function of the evolution
variables $k^2_\perp$ and $z$:
\begin{equation}
 P^P(k^2_\perp,z) = \frac{\alpha_s}{(2\pi)^2} \frac{1}{k^2_\perp}P(z)\;,\label{eq:pythia_split}
\end{equation}
where $P(z)$ is the appropriate unregulated Altarelli-Parisi splitting
function. There is an additional factor of
$1/(2\pi)$ to average over azimuthal angle, since the matrix elements
in \cref{eq:hej_split} will be evaluated for a given choice of
azimuthal angle for the generated emission. In the case of a space-like
branching we modify \cref{eq:pythia_split} to include the ratio
of PDFs for the branching, as in \cref{eq:hej_split_space}.

\begin{figure}[t]
\centering
\begin{subfigure}{0.495\linewidth}
        \includegraphics[width=\linewidth]{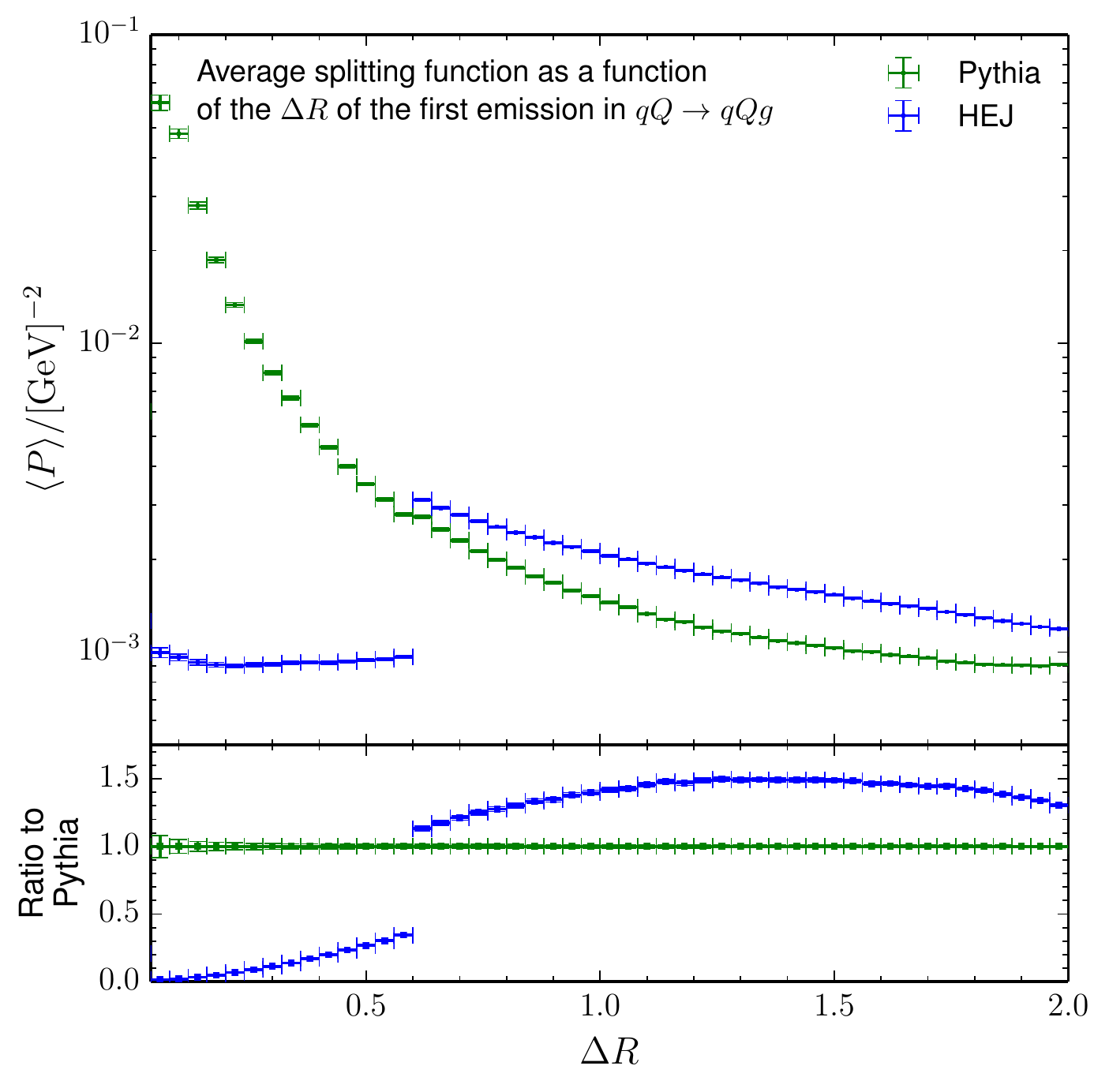}
    \caption{}
    \label{fig:splitDR}
  \end{subfigure}
  \begin{subfigure}{0.495\linewidth}
        \includegraphics[width=\linewidth]{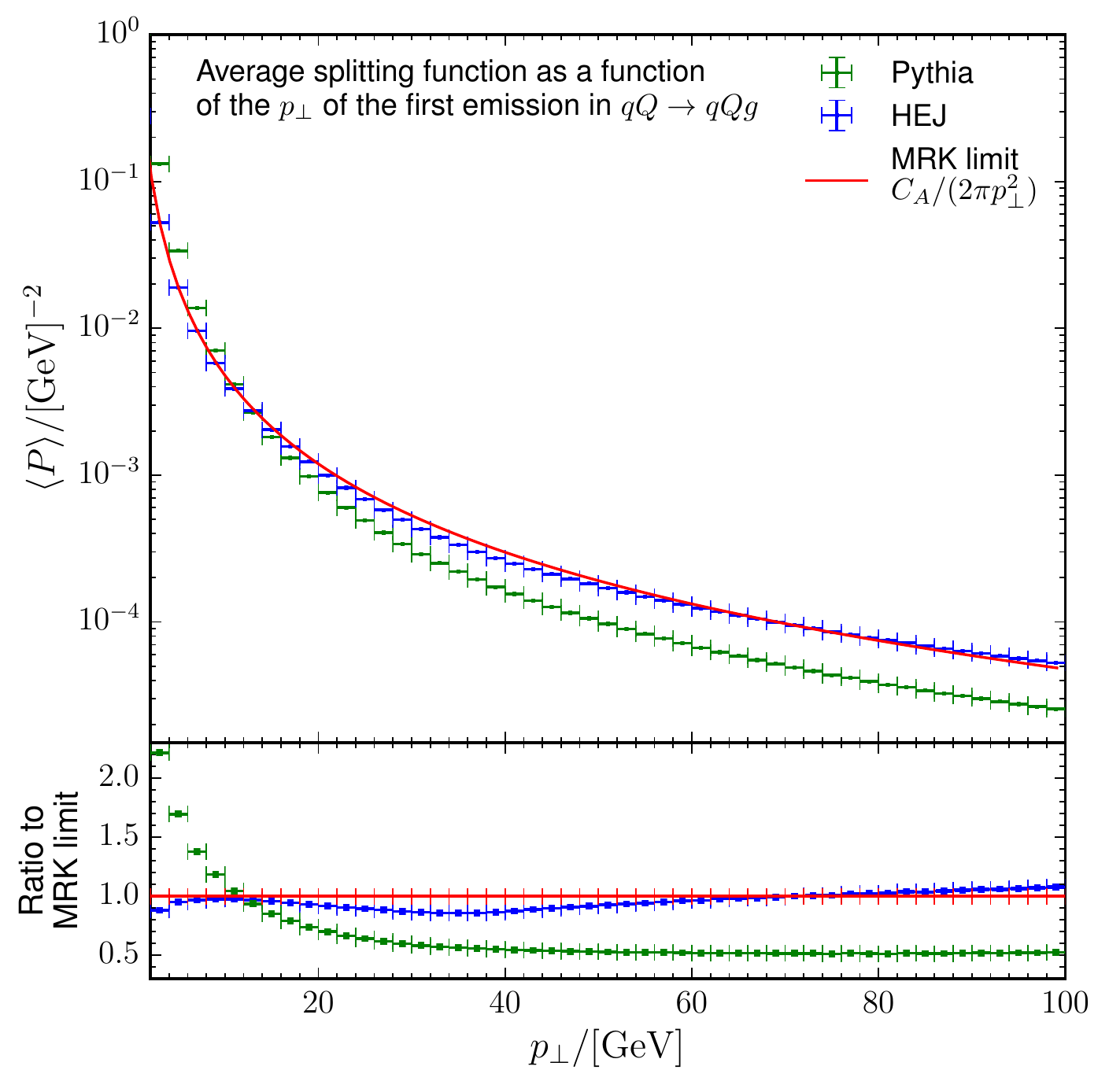}
    \caption{}
    \label{fig:splitkt}
  \end{subfigure}
  \caption{Plots comparing the average splitting function for \HEJ and \pyt for the first parton shower
  emission in \HEJ-generated $qQ\rightarrow qQ$ events.}
  \label{fig:splitPlots}
\end{figure}

 To illustrate the differences between the
  \HEJ and \pyt splitting functions, in \cref{fig:splitPlots} we show
  their typical behaviour as a function of (\subref{fig:splitDR}) the angular distance between the
  emitted gluon and the nearest parton, $\Delta R$, and (\subref{fig:splitkt}) the transverse
  momentum of the emitted parton in the lab frame, $p_\perp$. What is
  shown is the average value of the splitting functions in the first
  emission in \HEJ-generated $qQ\rightarrow qQ$ events, excluding
  factors of $\alpha_S$ and ratios of PDFs.
  In \cref{fig:splitDR} we average over emissions with $p_\perp>10$\GeV. 
  The
  discontinuity in the \HEJ splitting function at the jet radius is an
  artefact of the regularisation procedure used for emissions inside the
  jet cone of extremal jets \cite{Andersen:2011hs}. 
  This is in any case the
  phase space region we want to populate with collinear shower
  splittings.

For the $p_\perp$ plot we average of all
  emissions with $\Delta R>0.6$. We clearly see that for small
  $p_\perp$ the \pyt splitting function exceeds the \HEJ one (and
  also the analytic MRK-limit splitting shown for comparison). This
  is the region of large transverse momentum hierarchies, where the
  MRK approximation fails to properly resum the corresponding
  logarithms. Such large logarithms are present and are resummed by
  \pyt, and we would therefore like to add such splittings even if
  they are far away from the collinear region.

  From this we see that it makes sense to use a simple phase space cut
  based on the relative sizes of the splitting functions. 
  However, instead we can go one step further and in regions where 
  $P^P>P^H$ we subtract the \HEJ splitting function from the \pyt one.
  There we define the subtracted \pyt splitting function as

  \begin{equation}
\label{eq:subtracted_splitting_function}
P^S(k^2_\perp,z)= \max\left(P^P(k^2_\perp,z) -P^H(k^2_\perp,z), 0\right).
\end{equation}
%
%
Here the arguments of $P^H$ are intended to be schematic. This is
intended to denote that having generated an emission with
corresponding evolution variables $k^2_\perp$ and $z$, and having
inserted this into the event with an appropriate recoil strategy, the
matrix element containing $n+1$ partons should be evaluated with the
resulting set of $n+1$ final-state (recoiled) momenta. 
With this notation we can now define the subtracted Sudakov factor:
\begin{equation}
 \label{eq:modsudakov_merging}
 \Delta^S(\kti{i},\kti{i+1})= \exp \left\{ - \int_{\kti{i+1}}^{\kti{i}} d k_\perp^2  \int dz \Theta(P^P-P^H) 
   \left[P^P(k^2_\perp,z) - P^H(k^2_\perp,z) \right] \right\}	.
\end{equation}
It should be clear that we then want $\Delta^M = \Delta^S \Delta^H$,
and in order that such a Sudakov factor might be employed during a
trial shower, it is sufficient to generate emissions according to the
full \pyt splitting function $P^P(k^2_\perp,z)$, but veto emissions
with probability
\begin{equation}
 \mathcal{P}_\mathrm{veto} = P^H(k^2_\perp,z) /P^P(k^2_\perp,z),
\end{equation}
in accordance with the Sudakov veto algorithm.

Armed with this we can go through the steps in \cref{sec:ckkw-l-hej}
again and arrive at exactly the same formulae except with $P^C$ and
$\Delta^C$ replaced by $P^S$ and $\Delta^S$. The net result is that in
phase space regions where $P^H>P^P$, where we believe \HEJ is doing a
good job, we never add any \pyt splittings, while in the complementary
region emissions are added in proportion to the subtracted splitting
function so that in total they will correspond to populating that
region only with \pyt splittings.

\subsection{The Merging Algorithm}
\label{sec:ModifiedMergingAlgo}
For completeness we now explicitly disclose the full algorithm for merging \HEJ with \pyt as follows:
\begin{enumerate}
\item Generate samples of $n$-parton \HEJ states with $n \leq N$. Recluster any partons that have momenta above ${k_\perp}_{M}$ in such a way that the rapidities
of the resulting jets is unchanged. 

\item 
  For each $n$-parton state from \HEJ ($2<n\leq N$), reconstruct all possible \pyt
  shower histories where each clustering has the reconstructed scale $\kti{i}$, and set $\kti{n+1}$ $ = \kti{M}$. If $n=2$
  calculate the scale $\kti{2}$ and continue to step \ref{step:shower}, and otherwise continue as follows.
  \begin{enumerate}

  \item Throw away all histories that do not correspond to a sequence
    of \HEJ states.
  
  \item If there is at least one history that is correctly ordered in
    $\kti{i}$, throw away every other history.
  
  \item Give each history that is left a weight proportional
    to the \HEJ matrix element squared for the lowest multiplicity (\HEJ) state, times
    the product of \pyt splitting functions for the sequence of emissions 
    that gives the original $n$-parton state. Pick a history at random according to its relative weight.
    
  \item Starting from the most clustered state in the history, 
    make a trial emission from each intermediate state in the selected history
    starting from $\kti{i}$.

    \begin{enumerate}
     \item \label{step:trialemission} If the emission scale is below the reconstructed scale of the next state in the history, $\kti{i+1}$ ,
     continue to the next state in the history. If this is the original event we started with, 
     continue to step \ref{step:shower}.
     
     \item If the emission scale is above the reconstructed scale of the next state in the history,
     but has produced a \HEJ state, veto the emission with probability $P^H/P^P$.
     If the emission is vetoed, generate a new trial emission starting from the current
       emission scale, and return to \ref{step:trialemission}.
       If the emission is not vetoed replace the original event with this state 
       and continue to step \ref{step:shower}.
    \item If the emission scale is above the reconstructed scale and has not produced a \HEJ state, we
       substitute the original event with this state and continue to step \ref{step:shower}.
    \end{enumerate}  
  
  \end{enumerate}
  \item \label{step:shower}
  \begin{enumerate}
   \item If in the previous step we replaced the original event with one that could not have been produced by \HEJ,
    continue the shower from the emission scale of the new state without restriction.
   \item If this is the original event and we have $n < N$ start the shower from the reconstructed scale $\kti{n}$ and check the first emission. 
   If it gives a new  \HEJ state, discard the emission with probability $P^H/P^P$ and continue generating the first emission starting from 
   the scale $\kti{n}$.
   Once a first emission is accepted, the shower continues from the emission scale, radiating freely.
   \item If $n=N$, let the shower radiate freely from the scale $\kti{N}$.
  \end{enumerate}
  \item Once the parton shower has evolved below the cut-off scale, hadronise the event.
\end{enumerate}

This method represents one of the possibilities for merging \HEJ with a
parton shower. In particular, it retains the dijet cross section and
logarithmic accuracy of \HEJ: indeed, each event configuration and weight is
first generated by \HEJ, and all of phase space is thus covered. In the MRK
limits of similar transverse scales for all emissions, the Sudakov factors
introduced in equation~(\ref{eq:mergems}) all evaluate to unity, since
the scale used in the evolution of \pyt in the MRK limit tends to the
transverse scale of the lab frame. Since the MRK phase space is populated,
and the matrix elements are unchanged in this limit, the merging maintains the
logarithmic accuracy of \HEJ.

The logarithmic accuracy of \pyt is ensured since the full allowed phase
space in \pyt is covered, and the appropriate Sudakov factors between
emissions are applied in the shower evolution, with the possibility of
generating further emissions from the shower evolution. Such emissions are
then vetoed with a probability that said emissions were also generated from
\HEJ, such that double-counting is avoided.

For completeness we here mention three potential issues with the
algorithm. One limitation of the proposed method is that only the hardest
emissions (as ordered by \pyt) will be merged to \HEJ, which is not itself
ordered in hardness: it is possible for the parton shower to modify a state
classified as non-FKL (according to the momenta above the merging scale) to a
FKL state (accounted for in \HEJ) through an emission, and such emissions
will not have their splitting kernels subtracted. 
However, the non-FKL configurations account for a logarithmically suppressed
part of the cross section, quickly diminishing with increasing
rapidity \cite{Andersen:2017kfc}. Furthermore, future accounting for
next-to-leading logarithmic contributions in \HEJ will decrease further the
significance of the parton shower changing non-\HEJ to \HEJ states.

In addition, we reiterate that the method we are presenting is currently only applicable to FKL configurations. 
The impact of non-FKL corrections on the observables presented in this study is relatively small and within the indicated scale variations of the FKL results.
To include non-FKL configurations, it would be necessary to extend the definition of what constitutes a \HEJ state, and ensure that the appropriate tree-level matrix elements are used when calculating the veto probability for non-FKL states.
This is so that no problems arise from double-counting. Primarily such changes would affect what states may be included in the parton shower history, and which states may be inserted by the parton shower.

Finally we note that the factor one half in \cref{eq:hej_split} is based on
the fact that the colour flows which have a leading logarithmic
contribution will contribute equally to the colour-summed matrix element
squared in the MRK limit. This means that there will always be just two
possible colour flows for inserting a gluon in the exchange, and they will
have the same leading kinematic term in the MRK limit. While it is relevant
to take into account the different kinematic contribution from each possible
colour connection when matching full matrix elements to the parton shower,
the fact that the collinear divergences are absent from the formalism of \HEJ
means that the kinematic contributions from different colour connections
differ far less than in the full theory. While it would be possible to
account for the colour flow dependence in the contribution from the
sub-leading (and non-divergent) terms introduced in \HEJ compared to BFKL, we
choose in this study to assign an equal weight to the each of the possible
colour flows, just as will be the case in the MRK limit. This allows for the
simple attribution of $\frac 1 2 $ in \cref{eq:hej_split}.




\section{Results}
\label{sec:results}
\label{sec:descr-jet-struct}

\begin{figure}[t]
  \centering
   \includegraphics[width=.75\linewidth]{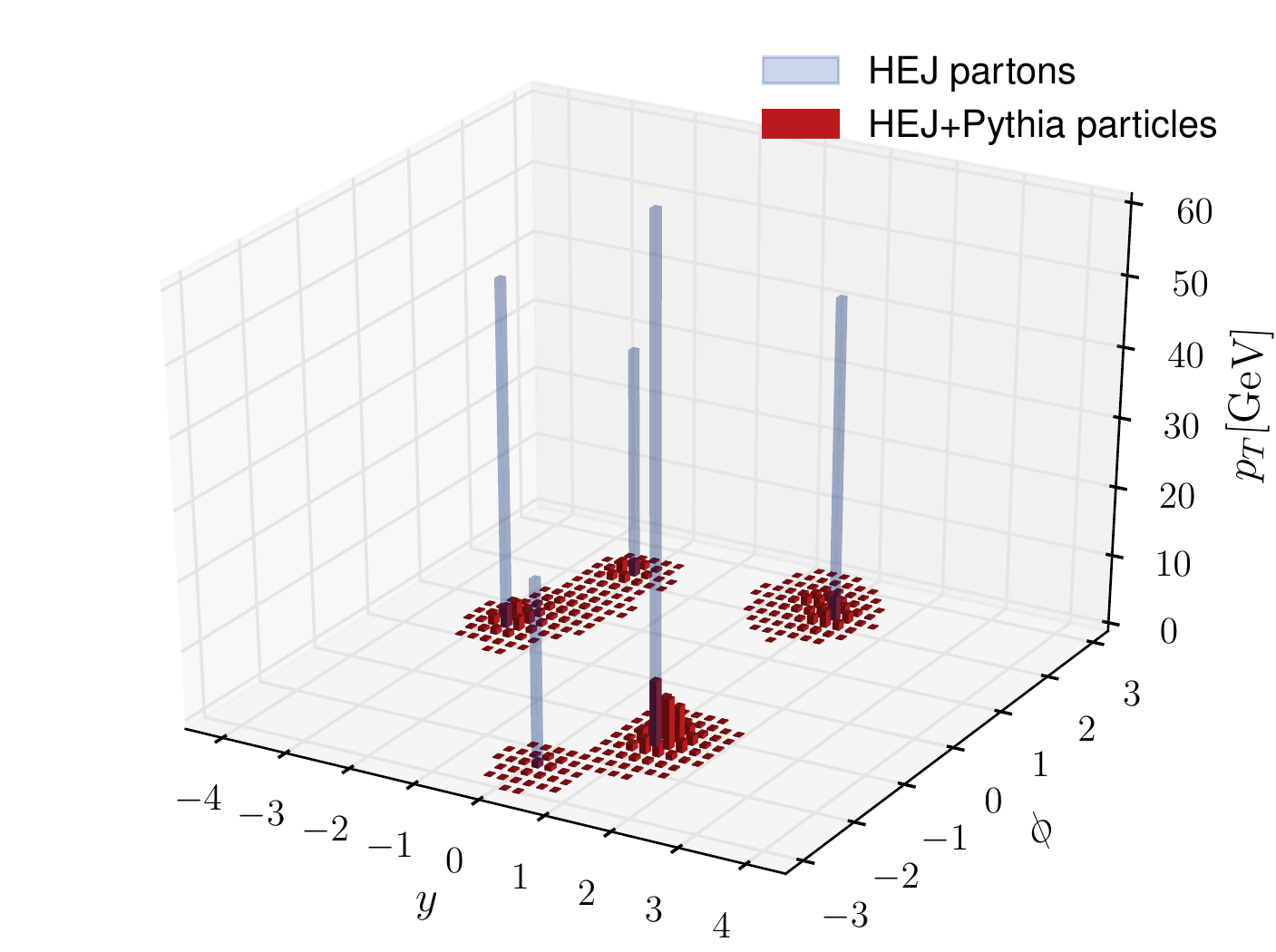}
  \caption{A Lego-plot of the momenta of partons arising from a single event
    from HEJ (blue) and average of the momenta arising from particles from
    that event showered 10,000 times with \py using the merging of
    \HEJpy. For this event configuration from \HEJ, which contains partons of
    similar transverse momenta, the effect of the showering is mostly to
    distribute physical particles around the partons of \HEJ.}
  \label{fig:lego}
\end{figure}

In this section, we will present the results of the formalism developed, and
contrast it with experimental data. We start however by qualitatively examining the
effect of the parton shower and hadronisation on a specific partonic event from \HEJ.
This is presented in \cref{fig:lego}, which shows a LEGO-plot of the average transverse momentum
deposit (greater than 100\MeV) in bins of $0.2\times 0.2$ units of rapidity and azimuthal angle. 
A single \HEJ event with 5 partons (and with fixed colour connections),
of which 4 are sufficiently hard to form individual jets of
$p_T>30$\GeV, is shown in blue. The average result of passing this event 10,000 times to \py
is shown in red. 

The effect of the shower on average is to spread out
the momentum of each \HEJ parton over an area with radius
$R\sim 1$ around that parton.
Indeed, for events similar to the chosen one, the effect of
the shower seems to be limited to filling the jet cones, and in \cref{sec:jetprofiles}
we study in more detail the accuracy with which the jet
cones are filled. In \cref{sec:impact-multi-jet} we will study
multi-jet observables, some of which probe large hierarchies in transverse momentum, and in such regions 
\py can additionally supplement the jets produced by \HEJ. 

We will mainly look at LHC analyses especially designed to probe
effects of high energy logarithms. It should be noted however, that
these have so far employed a relatively soft definition of
hadronic jets, typically requiring a transverse momentum of less than
40\GeV. This results in broad shower profiles, where the description
of the spill-over outside the jet cones is necessary for accurate
results. Furthermore, it was noted in ref.~\cite{Alioli:2012tp} that
these analyses often use cuts that also enhance soft and collinear
logarithms. In \cref{sec:impact-multi-jet} we therefore
propose to use a slightly harder threshold for jets to reduce the dependence
on shower and hadronisation effects, and crucially, investigate the full
rapidity range of the hard event rather than just the region in-between the
two hardest jets. This allows for a much
cleaner probe of the high energy logarithms.


We note that there exist many parameters in \py that 
control non-perturbative effects, and which are fixed by 
tuning to measurements of certain soft observables. We investigate an example of such an observable in the next section. 
As we will shortly see, the combination of \HEJ and \py obtains a very similar description
to \py alone. Therefore, in the results that follow 
we do not retune \py for use with \HEJ, 
even if this might further improve the agreement of
\HEJpy with data; instead we use the default Monash 2013 tune \cite{Skands:2014pea}
for both \py and \HEJpy.

\subsection{The Description of the Profile of Jets}
\label{sec:jetprofiles}

\begin{figure}[thp]
  \centering
     \includegraphics[width=\linewidth]{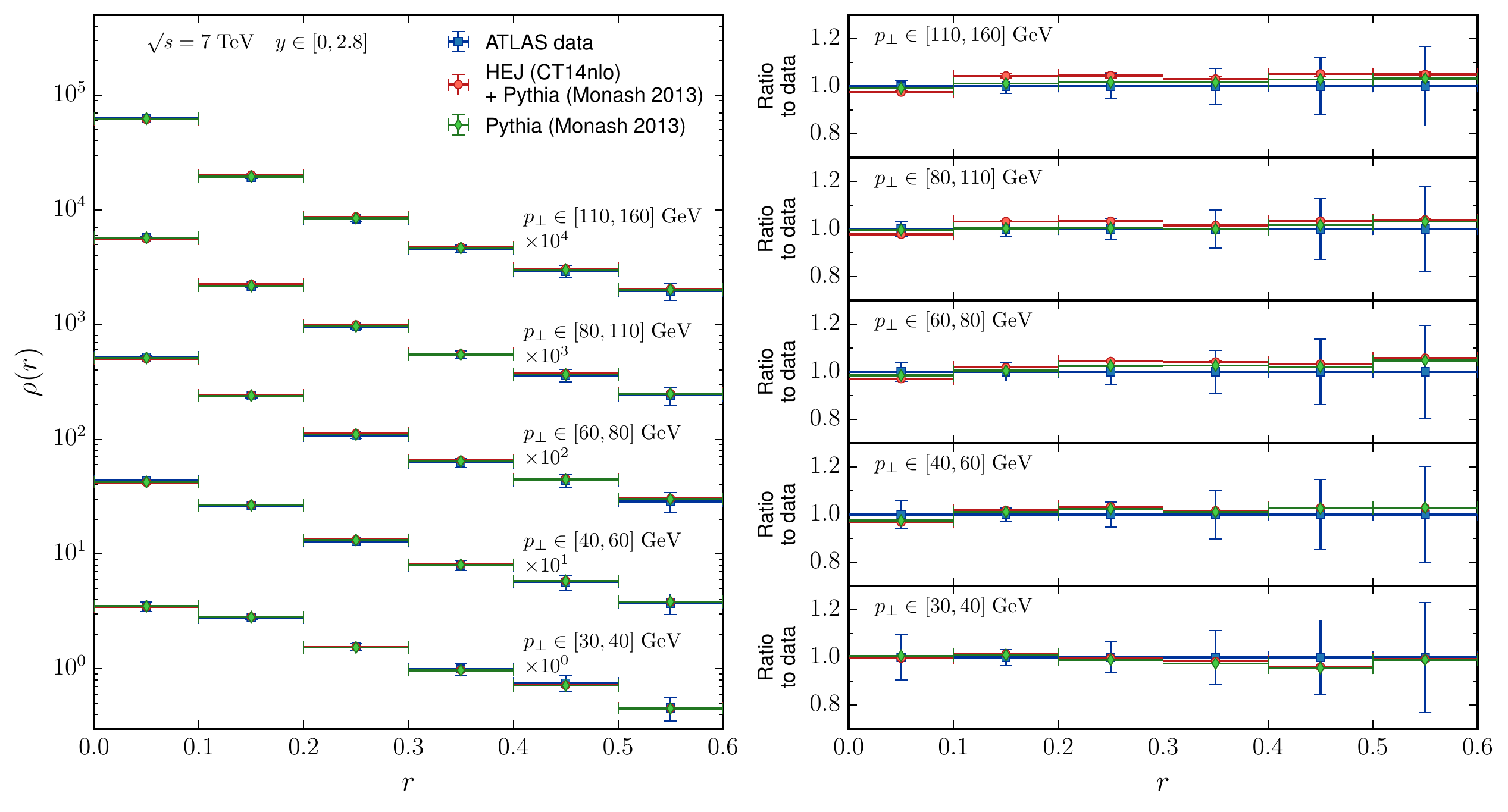}
  \caption{The data and predictions for the differential jet profile as
    defined in \cref{eq:rhoofr}. The parton shower of \py gives a very good
    description of data, which is inherited by \HEJpy.}
  \label{fig:jetprofile}
\end{figure}

The \emph{jet profiles} were measured at the LHC 
in early 7\TeV\ runs, for example by ATLAS in ref.\@ \cite{Aad:2011kq},  
accepting events with just one primary vertex (no pile-up) and at least one
jet with transverse momentum $p_\perp>30$\GeV\ and rapidity $|y|<2.8$. For such events, the differential
jet profile $\rho(r)$ as a function of the distance
$r=\sqrt{\Delta y^2+\Delta \phi^2}$ to the jet axis is defined as the average
fraction of the jet transverse momentum in an annulus between $r-\Delta r/2$
and $r+\Delta r/2$ around the jet axis in the $(y,\phi)$-plane. As such, 
\begin{equation}
  \label{eq:rhoofr}
  \rho(r)=\frac 1 {\Delta r} \frac 1
  {N_{\mathrm{jets}}}\sum_{\mathrm{jets}}\frac{p_\perp(r-\Delta r/2,r+\Delta r/2)}{p_\perp(0,R)},
\end{equation}
where $p_\perp(r_1,r_2)$ is the summed $p_\perp$ in the annulus of the two
circles of radii $r_1$ and $r_2$, and $N_{\mathrm{jets}}$ is the number of
jets. The measurement of ref.\@ \cite{Aad:2011kq} used $\Delta r=0.1$ and
the anti-$k_T$ jet algorithm \cite{Cacciari:2008gp} with $R=0.6$. This
analysis is implemented in recent versions of Rivet \cite{Buckley:2010ar},
which we use to analyse generated events (both here and in \cref{sec:impact-multi-jet}).

The \HEJ formalism captures the logarithms associated with wide-angle
emissions, but not those associated with the collinear emissions. \HEJ is thus
not expected to fill the jet cones with radiation, and it is expected that
the results of \HEJ\!+\py for the jet shapes is similar to those of pure
\py (since the merging procedure should produce results similar to \py in
regions where \HEJ does not radiate). In \cref{fig:jetprofile} we compare 
the predictions of \py and \HEJpy for $\rho(r)$ in slices of jet transverse momentum
to data \cite{Aad:2011kq}. While
\HEJ alone would primarily have filled just the first bin in each distribution,
\HEJpy gives the same very good description of the jet shapes
as the parton shower of \py. The merging procedure has therefore performed a
perfect job of populating the jet areas (through collinear emissions), which
are mostly empty in the pure partonic description of \HEJ --- and has of
course furthermore fully hadronised the partonic states.
The ability to describe this observable represents an improvement relative to
the matching of \HEJ + \ariadne.

\subsection{Impact on Multi-jet Observables}
\label{sec:impact-multi-jet}

In this section we will investigate the impact of the merging on
observables which depend only on the identified hard jets of the event. 
We shall make comparisons between pure \HEJ, \py and \HEJpy. Events in \HEJ 
(both with and without showering) were
generated with the PDF set CT14nlo \cite{Dulat:2015mca,Buckley:2014ana}, and 
the renormalisation and factorisation scales were taken to 
be the maximum jet transverse momentum $\mu_R=\mu_F = {p_T}_\mathrm{max}$.
In the case of pure \HEJ the scale uncertainties were estimated by varying 
$\mu_R$ and $\mu_F$ independently between twice and half the central scale choice,
and these uncertainties will be denoted as a band around the central predictions.
The vertical lines indicate the statistical uncertainty on the results. 
As before, \py predictions were generated using the Monash 2013 tune. The
expectation is that there should be little impact on the results of \HEJ in
phase space regions where the jets have similar transverse momenta but are widely
separated in rapidity. This is the region where \HEJ should already control
the dominant logarithms to all orders. The parton shower should therefore not
introduce sizeable corrections. On the other hand, as mentioned in \cref{sec:hejformalism}
regions with large disparate transverse scales are not encompassed by the kinematic assumptions
of the \HEJ formalism, and should therefore receive additional hard emissions from the parton
shower. 

\begin{figure}[t]
  \centering
     \includegraphics[width=\linewidth]{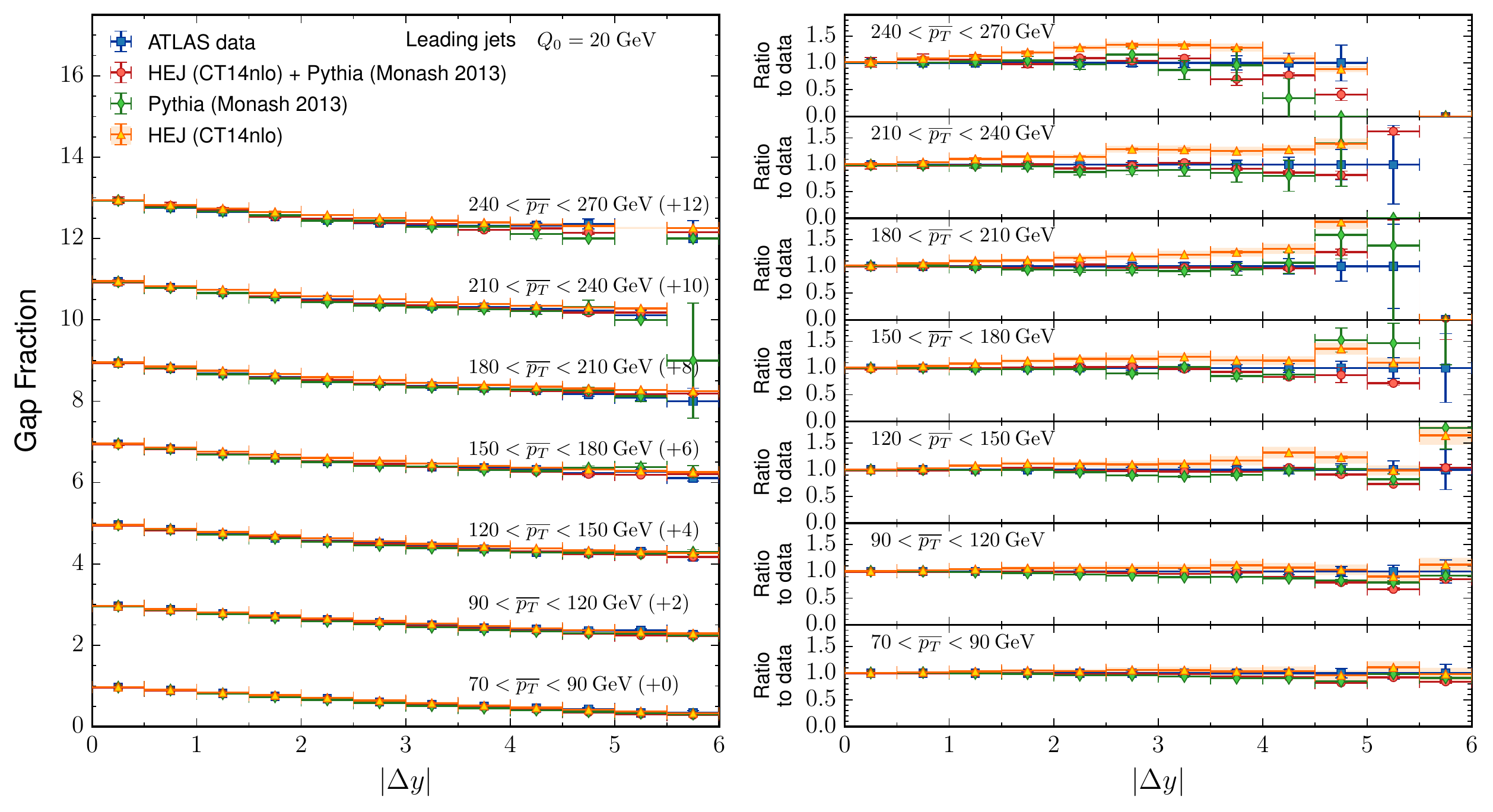}
  \caption{Plot showing a comparison between \HEJ, \py, \HEJpy  and ATLAS data \cite{Aad:2011jz} for
  the gap fraction as a function of the rapidity separation of the tagging jets, in slices of the average transverse momentum of the tagging dijets.
  The impact of the parton shower in \HEJpy is modest.}
  \label{fig:gapfrac}
\end{figure}

We will first consider two ATLAS analyses \cite{Aad:2011jz,Aad:2014pua}
that measure the amount of additional radiation in inclusive dijet events. 
Dijet systems are of course simple
at the Born level, characterised by two jets of equal transverse momenta that are
back-to-back in the azimuthal plane; however, this simple topology is in
general spoiled by radiative corrections. 
The analyses in question both require the existence of a dijet pair above 
some transverse momentum cut, defining the tagging jets; in what follows 
the tagging jets are identified as the two hardest (leading) jets in the event.
The number of jets in the rapidity interval between the tagging 
jets, each having a transverse momentum above a given \textit{veto scale} $Q_0$,
is then measured. 
This allows the definition of two observables: first, the \textit{gap fraction}, 
and secondly the average number of jets in the rapidity interval $\overline{N_\mathrm{jet}}$.
Events having no jets above the veto scale in the rapidity interval between the tagging jets
are classified as gap events. The gap fraction as defined by ATLAS \cite{Aad:2011jz,Aad:2014pua}
is then simply the ratio of the contribution to the cross section from these gap events to 
the inclusive dijet cross section.


\begin{figure}[t]
  \centering
   \includegraphics[width=\linewidth]{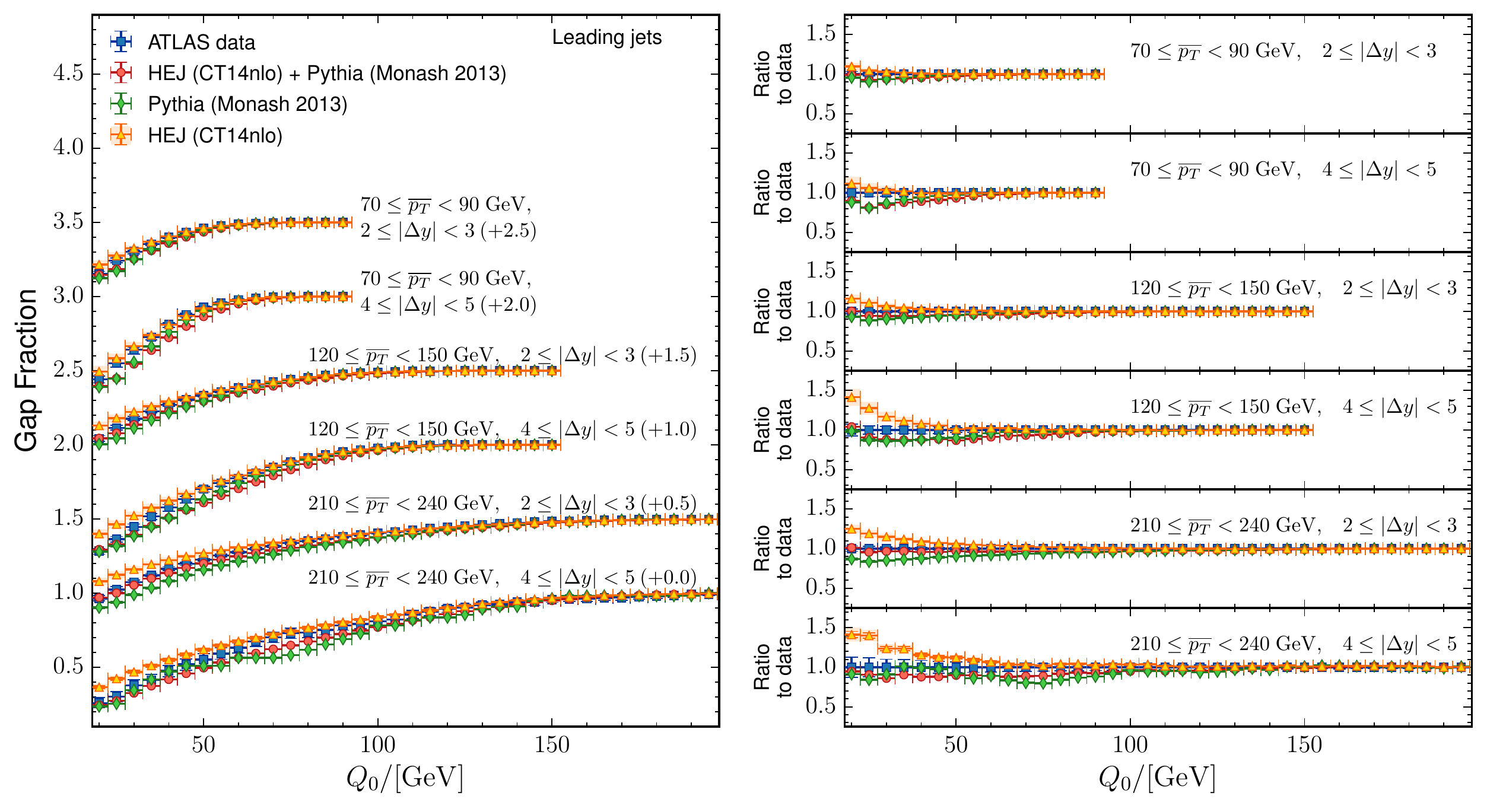}
  \caption{Plot showing a comparison between \HEJ, \py, \HEJpy  and ATLAS data \cite{Aad:2011jz} for
  the gap fraction as a function of the veto scale $Q_0$, in slices of both the average transverse momentum and rapidity separation 
  of the tagging dijets.  For sufficiently large $Q_0\sim 50$\GeV\ HEJ alone
    achieving a good description; \py and \HEJpy are consistent, with a good description across all bins. 
    }
    \label{fig:gapfracb}
  \label{fig:gapfrac2}
\end{figure}

We start with the ATLAS analysis presented in ref.\@  \cite{Aad:2011jz}, 
in which jets were defined using the anti-$k_T$ jet
algorithm with $R=0.6$ and having rapidity $|y_j|<4.4$.
In \cref{fig:gapfrac} we show a plot of gap fraction as a function of the rapidity interval between the tagging jets
$|\Delta y|$,
where the veto scale was taken to be $Q_0 = 20$\GeV.
This is shown in bins of the average transverse momentum of the tagging dijets $\overline{p_T}$,
from  70\GeV\ -- 90\GeV\ to 240\GeV\ -- 270\GeV.
By construction,
the gap fraction will be 1 at $|\Delta y|=0$, since the phase space where a
third jet would be counted is vanishing (since only jets in-between the two
hardest jets are counted, and the rapidity difference between the two
hardest jets is zero). 
The variation between the predictions is small, and
discernible only for the bins with the largest $\overline{p_T}$.
Here there is a large hierarchy between the scale of the tagging jets
and the scale of any additional jets (which are characterised by the veto scale).
It is therefore not surprising that \HEJ predicts too few additional jets
in this region instead requiring the DGLAP resummation of 
the parton shower. Moreover the combination of \HEJpy results in a description 
that at least as good as, or better than, \py or \HEJ
individually.

In \cref{fig:gapfrac2} the gap fraction is instead shown as a function of the 
veto scale $Q_0$, now binned in both $\overline{p_T}$ and $|\Delta y|$.
It is evident that even a modest increase in $Q_0$ to 50\GeV\ in \cref{fig:gapfrac}
would have brought the predictions
from pure \HEJ into perfect agreement with data across all regions in
$\overline{p_T}	$ and $\Delta y$. Furthermore, there are indications (e.g.~from the
region of
$210\GeV\le\overline{p_T}\le240\GeV, 2\le\Delta y\le3$) that
the high energy logarithms of \HEJ in \HEJpy improve the predictions of \py
alone.

\begin{figure}[t]
  \centering
  \begin{subfigure}[t]{0.495\linewidth}
   \includegraphics[width=\linewidth]{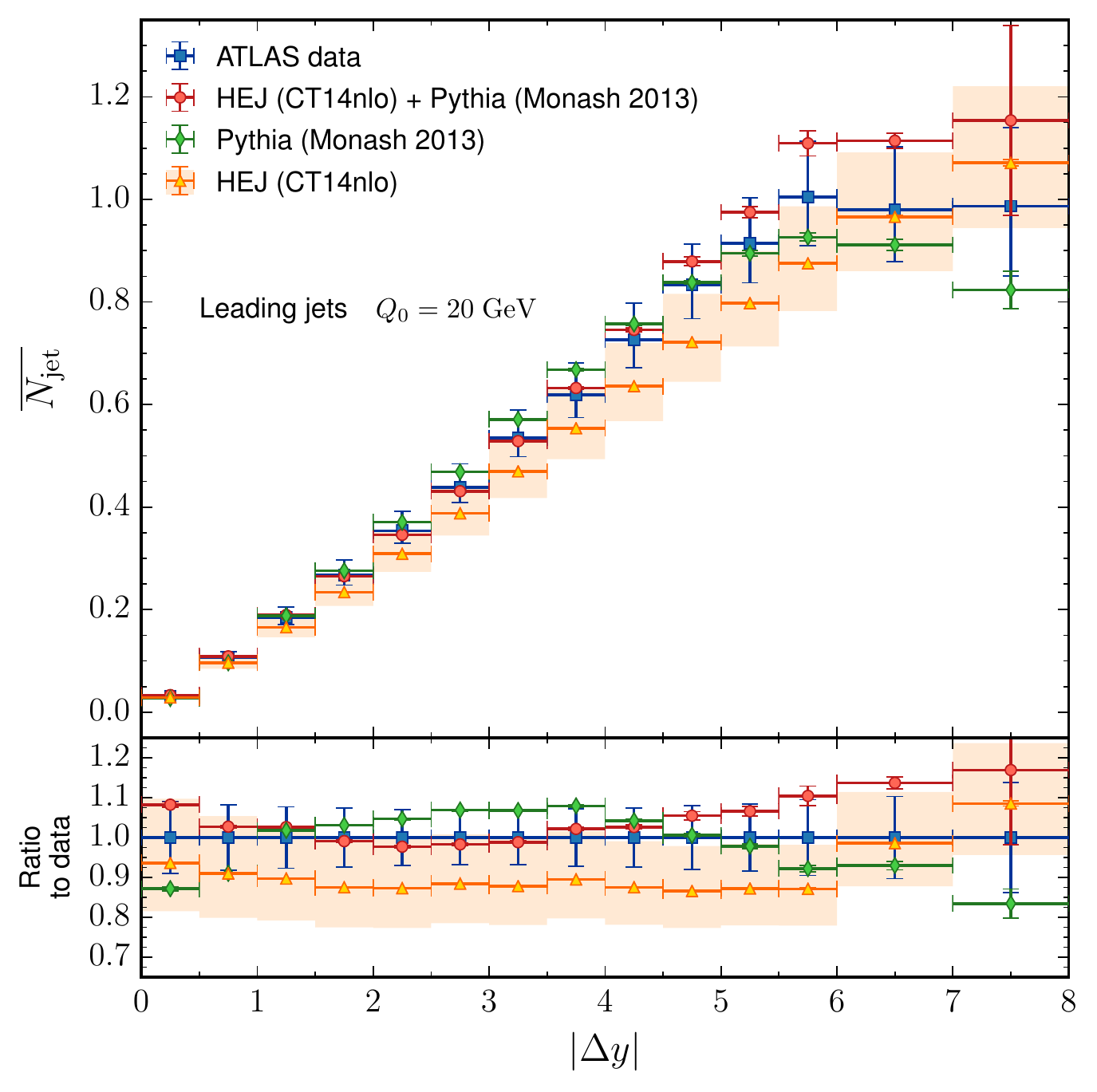}
    \caption{}
    \label{fig:avgjets20GeVHardestDY}
  \end{subfigure}
  \begin{subfigure}[t]{0.495\linewidth}
\includegraphics[width=\linewidth]{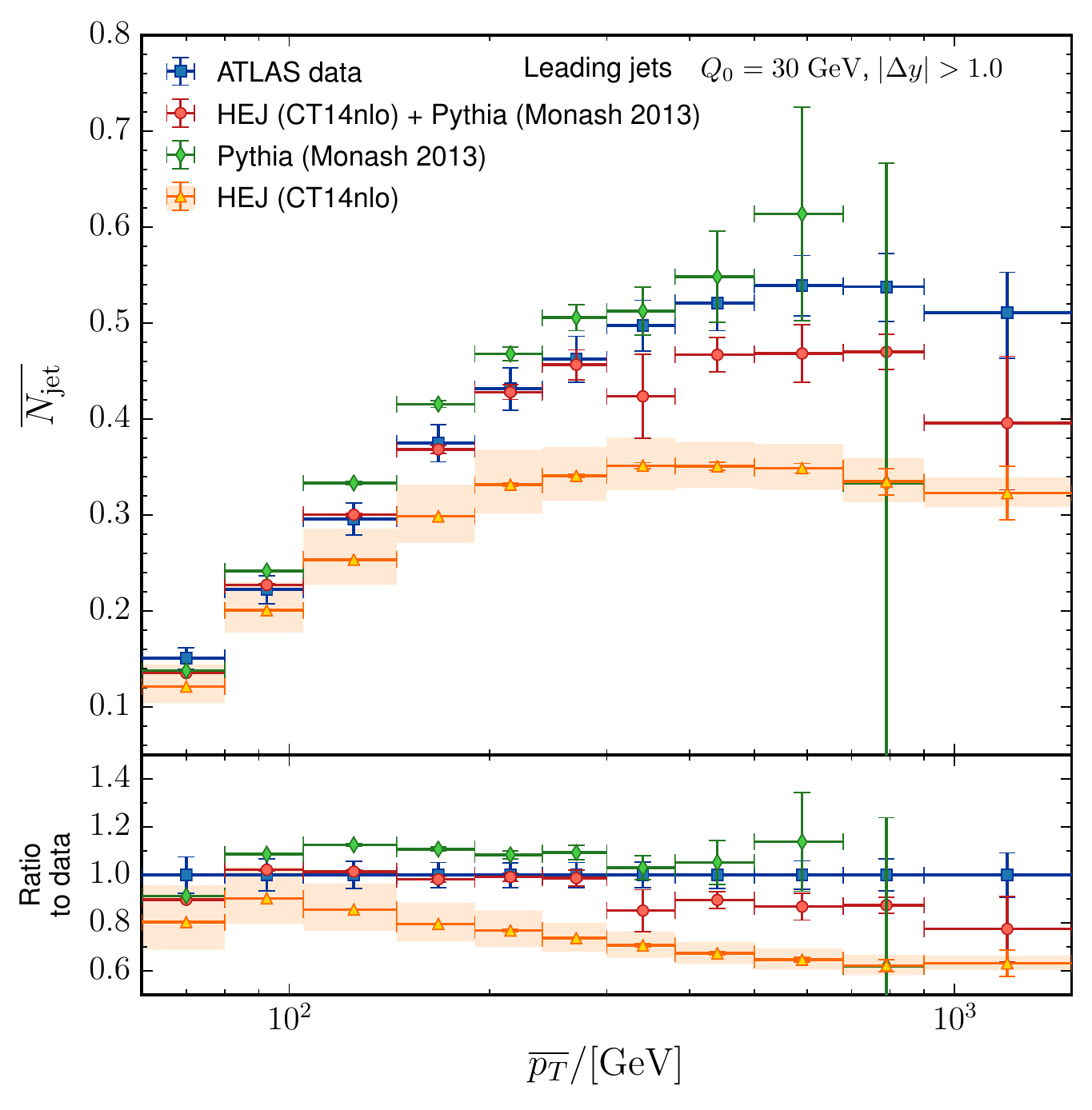}
    \caption{}
    \label{fig:avgjets20GeVHardestPT}
  \end{subfigure}
  \caption{Plot showing a comparison between \HEJ, \py, \HEJpy  and ATLAS data \cite{Aad:2014pua} for
  the average number of jets in the rapidity interval between the two tagging jets, as a
    function of  (\subref{fig:avgjets20GeVHardestDY}) the rapidity interval
    between the two tagging jets, and (\subref{fig:avgjets20GeVHardestPT}) the
    average transverse momentum of the two tagging jets.}
    \label{fig:avgjets20GeVHardest}
\end{figure}

The average number of hard jets is a potentially better discriminant between
the predictions than the gap fraction, simply because the average number of
jets has a larger range of variation. We now consider this observable as measured
by ATLAS in ref.\@  \cite{Aad:2014pua}, where the hardest and second hardest jets 
(also defining the tagging jets)
were required to have transverse momenta above 60\GeV\ and 50\GeV\ respectively\footnote{
Asymmetric cuts are required in order for a meaningful
  comparison to NLO calculations, which suffers a spurious logarithmic
  dependence on the soft
  emissions \cite{Frixione:1997ks}.}.
Jets were again defined using the anti-$k_T$ algorithm with $R=0.6$.
In \cref{fig:avgjets20GeVHardestDY} the average number of jets in the interval between
the tagging jets is shown as a function of 
the rapidity interval between the tagging jets (with $Q_0 = 20$\GeV).
While the differences in the predictions are again small, we observe
that although the data from ATLAS lie within the scale uncertainty band for 
pure \HEJ, the central line for \HEJ nevertheless underestimates the 
number of additional jets. Meanwhile, the predictions of \HEJpy are better in line with data, and
are of a similar quality to that of \py. 

In \cref{fig:avgjets20GeVHardestPT} the average number of jets in the interval between
the tagging jets is instead shown as a function of the average transverse
momentum of the tagging jets (with $Q_0 = 30$\GeV), and where the dijets were
required to be separated by at least one unit of rapidity.
 As the average transverse momentum of the two
hardest jets increases to 1\TeV, the number of 30\GeV\ jets is unsurprisingly no
longer well-described without a shower resummation. Indeed, for increasing
$\overline{p_T}$, the predictions of pure \HEJ rises from 0.12 additional jets
to 0.3, whereas data rises from 0.15 to 0.5. Both \py and \HEJpy give a good
description of this distribution. For such large ratios of transverse scales,
the effect of the shower resummation is large, and therefore the results for 
\HEJpy are outside the scale uncertainty band for pure \HEJ.

It should be apparent at this stage that in distributions that probe large differences in transverse momentum such as 
\cref{fig:avgjets20GeVHardestPT}  a parton shower is necessary for an accurate description, and therefore the
addition of \py to \HEJ gives rise to a notable improvement relative to \HEJ. 
Likewise, in distributions that probe large rapidity spans, one might have expected that \HEJ (and hence \HEJpy)
would provide a superior description relative to \py. Indeed, it is perhaps surprising that the predictions of \HEJ and \py are so
similar for the rapidity distributions studied so far. 
(In fact, we note that in some cases the description of \py is closer to data
than the predictions of \pyt+\texttt{POWHEG}
\cite{Nason:2004rx,Frixione:2007vw,Alioli:2010xa} which were used in the
original analyses \cite{Aad:2011jz,Aad:2014pua}. This could be an effect of
the later tunings of the non-perturbative parameters of \pyt, and reiterates
the possible benefits of performing similar analysis with much harder jet
scales, such that the sensitivity to the tunings of the MPI and
non-perturbative effects are reduced).
Firstly, the restrictive definition of the chosen observables prevents much variation in 
their values. Also, as discussed in ref.\@  \cite{Alioli:2012tp}, the softness of the
veto scale relative to the tagging dijets' transverse momentum results in 
event samples that are influenced by \textit{both} high energy and soft-collinear logarithms,
spoiling the applicability of the \HEJ formalism. 

\begin{figure}[t]
  \centering
  \begin{subfigure}[t]{0.495\linewidth}
  \centering
   \includegraphics[width=\linewidth]{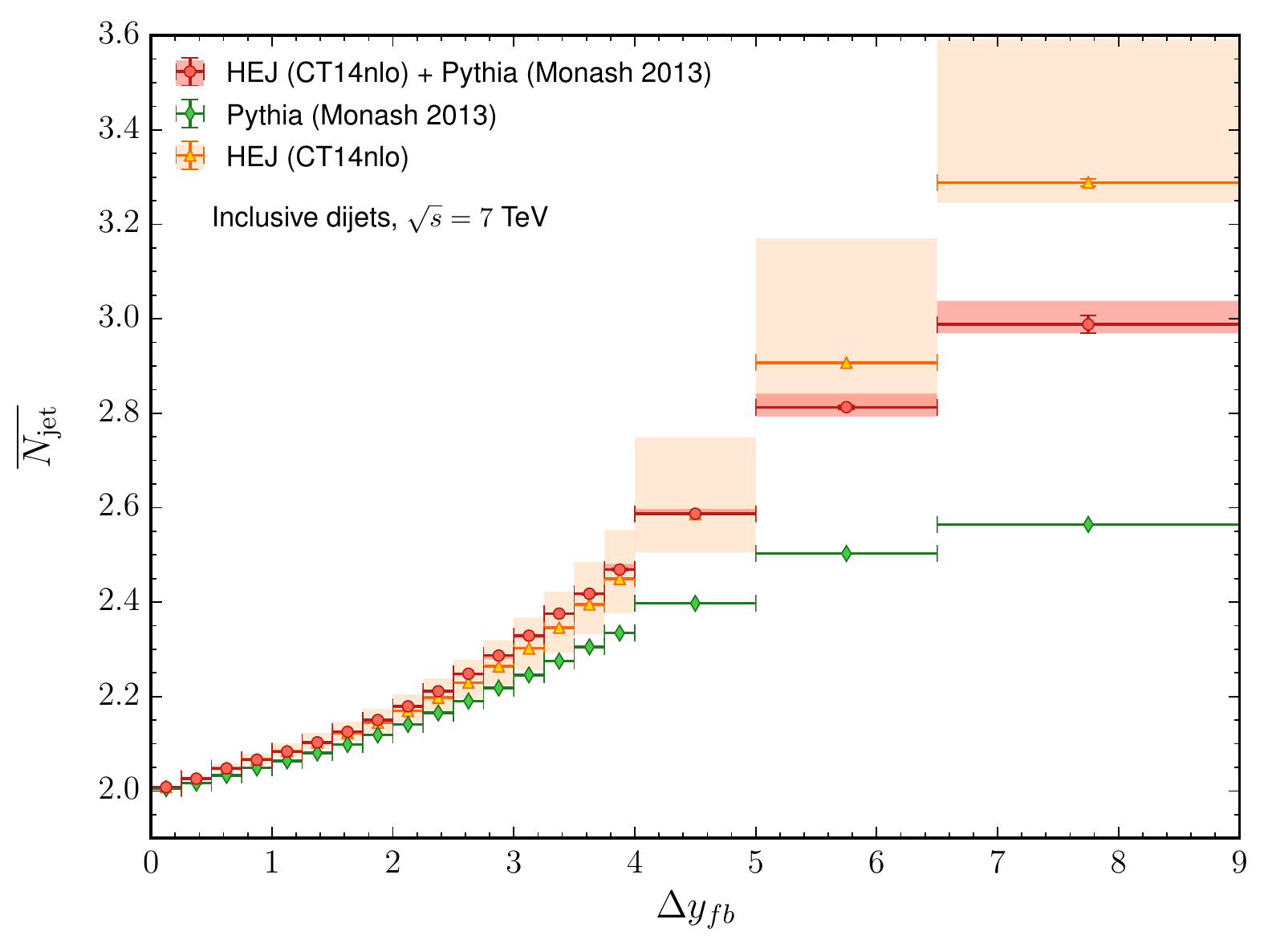}
  \caption{}
  \label{fig:avejetspure_dyfb}
\end{subfigure}
  \begin{subfigure}[t]{0.495\linewidth}
    \centering
     \includegraphics[width=\linewidth]{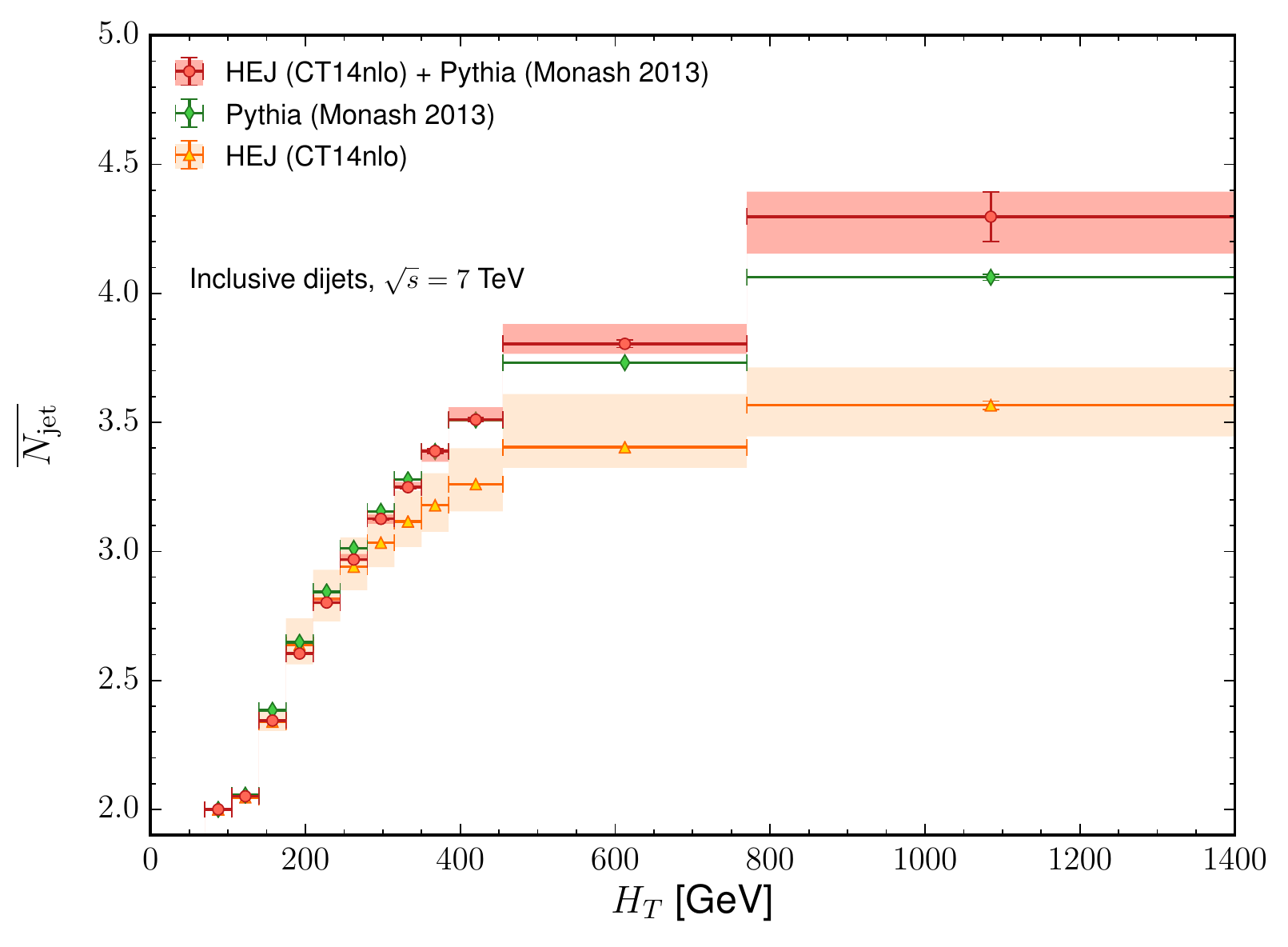}
    \caption{}
    \label{fig:avejetspure_HT}
\end{subfigure}  
\caption{Plot showing a comparison between \HEJ, \py and \HEJpy
for the average number of jets as function of (\subref{fig:avejetspure_dyfb}) 
the rapidity interval between the most forward and backward jets, and (\subref{fig:avejetspure_HT})
the scalar sum of transverse momenta. The event selection and definition of observables was taken 
from ref.\@  \cite{Alioli:2012tp}, chosen to better disentangle effects originating from high energy or soft and collinear logarithms.}
  \label{fig:avejetspure}
\end{figure}

Simpler event samples were
suggested in ref.\@ \cite{Alioli:2012tp} to disentangle the two sources of
logarithmic corrections, together with more inclusive observables that better expose the
differences in the description of a fixed-order calculation, a parton shower
and \HEJ. The analysis considered inclusive dijet events, requiring at least one jet with 
transverse momentum above 45\GeV, and with all other jets required to have transverse momentum above 35\GeV. 
Jets are defined using the anti-$k_T$ algorithm with $R=0.5$ and with rapidities $|y_j|<4.7$.
Comparisons between \py, \HEJ, and \HEJpy were made for this analysis
and the results for the average total number of jets are shown in \cref{fig:avejetspure}.
We emphasise that the additional jets are no longer required to be in the rapidity interval
between the two hardest jets, and there is no longer a significant disparity between their transverse
momenta and that of the two hardest jets. This results in a greater number of jets 
passing the selection cuts, and consequently the potential for variation between 
different predictions is slightly higher.

Also shown in \cref{fig:avejetspure} as a shaded red band around the central
predictions for \HEJpy are variations of the merging scale ${k_\perp}_{M}$
(with a central scale of 15\GeV) between 7.5\GeV and 30\GeV. ${k_\perp}_{M}$
should be set to a value below the minimum jet transverse momentum used in
the analysis, which in this case is 35\GeV.  We see that even for these very
exclusive multiplicity-dependent observables, allowing the merging scale to
get very close to the analysis scale leads to only modest variations, and do
not exceed the size of the \HEJ renormalisation and factorisation scale
uncertainties.  As this plot is most sensitive to differences between \HEJ,
\py and \HEJpy, we expect the merging scale dependence in other plots to be
comparable or smaller than that observed here.


In \cref{fig:avejetspure_dyfb} the average number of jets is shown as a function 
of the rapidity interval between the most forward and backward jets $\Delta y_{fb}$;
we expect this to be particularly sensitive to the logarithms in $\hat s/|\hat t| \sim e^{\Delta y}$
summed by \HEJ.
The predictions of \py are significantly lower than those of \HEJ and
\HEJpy, and moreover are outside the scale variation band for pure
\HEJ beyond $\Delta y_{fb}>4.5$. This implies that in this regime, the 
merging of \HEJ with \py increases the number of wide-angle jets relative to \py alone,
as we should expect.  Such differences should be even more pronounced
with a larger centre-of-mass energy than the choice of $\sqrt{s} =7$\TeV which was used for these
comparisons.

It is interesting to note that not only is the spread of predictions 
significantly larger in \cref{fig:avejetspure_dyfb} than in \cref{fig:avgjets20GeVHardestDY},
but also that the prediction of \HEJpy is
\emph{lower} than that of \HEJ alone.
There are several possible explanations for this. 
Firstly the addition of a parton shower extends the shower profile beyond the
jet radius, such that potentially fewer of the jets from the partonic calculation
pass the relevant criteria.
Secondly, at $\Delta y=10$ two partonic
jets of 45\GeV\ transverse momentum would take up all the energy available at
a 7\TeV\ collider, and all predictions for the average number of jets would
therefore have to return to 2 at this point. Since the parton shower uses some of
the available collider energy in (for example) the description of the underlying event,
the turnover of the average number of jets will have to happen
earlier than in the pure partonic prediction.
Alternatively it could be that too many non-FKL configurations of lower multiplicity are being inserted by
the merging algorithm, an issue that could be resolved by the extension of this method
to merge non-FKL events.

Finally, in \cref{fig:avejetspure_HT} the average number of jets is shown
as a function of the scalar sum of transverse momentum $H_T$; we expect this observable to 
be sensitive to the double logarithms in transverse momentum summed by the parton shower.
\py now adds further hard radiation to that of \HEJ, which is both as expected 
and is consistent with the previous results. 

The choice of more inclusive observables and simpler selection cuts leads to a clearer
separation of the effects of the logarithms included in the parton shower and
those of the all-order summation of high energy logarithms in BFKL or \HEJ.
A simple experimental investigation with a similar set of cuts and distributions 
would be extremely interesting in exposing
the shortcomings of either predictions, and the benefits of the combined
formalism presented in this paper. Such an experimental analysis would
further aid the development of predictions valid for the separation of the
VBF and GF contribution to Higgs-boson production in association with dijets.



\section{Summary and Outlook}
\label{sec:outlook}

We have introduced a new CKKW-L-inspired merging algorithm for combining the
all-order summation of high energy logarithms in \HEJ with a parton shower.  
For the first time
\HEJ events have been fully evolved down to particle level using the modern
parton shower, hadronisation and modelling of MPI in \py. The merging
algorithm systematically combines the dominant perturbative
corrections due to hierarchies of transverse momenta (i.e.~of soft and
collinear origin) from the parton shower with those due to large invariant
masses between jets of similar transverse momenta, as implemented in \HEJ or
BFKL.

The performance of the merging algorithm was assessed by considering observables which measure the additional radiation in the rapidity interval between two tagging jets. 
Many of the observables measured so far have (intentionally or not) a hierarchy of transverse scales
induced, and so require a systematic resummation of the
logarithms from the parton shower in order to arrive at a satisfactory
description. For such observables we find that 
the description of \HEJpy is consistent with standalone \pyt and data.
The improvement upon \HEJ in such distributions is notable. 
In addition, an investigation of related observables but with more inclusive cuts demonstrated that 
the jet multiplicity for large rapidity intervals is increased relative to
\pyt through merging. A measurement of such clean observables can serve as
test of high energy evolution.
These results demonstrate that we have combined effects originating from both parton shower and from \HEJ, providing a proof of concept for this method.


Notwithstanding what has been so far achieved, what has been presented constitutes a first attempt at merging \HEJ with a parton shower. We envisage numerous 
refinements that can be made. There is a need to implement a prescription for
incorporating full fixed-order matching into the merging procedure and the
inclusions of sub-leading partonic channels (non-FKL) to the \HEJ resummation\cite{Andersen:2017kfc}. 
In particular this will have an impact upon which states may be inserted by the parton
shower. A systematic
inclusion of such events in the prescription would not require dramatic changes to the algorithm. Firstly, the definition of what constitutes a \HEJ state would need to be extended;
secondly, the appropriate tree-level matrix elements should be used when calculating the veto probability of trial emissions. 
Nevertheless, the impact upon the observables discussed in this
paper should be relatively modest; this was assessed by studying the relative size of the 
contributions of fixed-order non-FKL events in pure \HEJ.

As discussed in \cref{sec:ModifiedMergingAlgo}, a limitation to the method is that only
the hardest emission of the shower received subtractions in their associated splitting kernel.
This limitation could be addressed by re-inserting \HEJ emissions in those events that were modified by 
\pyt above the merging scale, at the appropriate evolution scales reconstructed during the parton shower history.
However, such a procedure has several ambiguities, such as where in the (modified) colour flow the emission should be
inserted, and precisely how the recoils should be performed. Preliminary studies indicate the effect of
reinserting \HEJ emissions has a small effect, even upon the most sensitive observables shown in \cref{fig:avejetspure}.
However, we postpone a systematic study of these effects to a future publication. 

Finally, also noted  in \cref{sec:ModifiedMergingAlgo}, a more advanced treatment for the weighting of colour flows in \HEJ events
that is informed by the parton shower may be necessary. The impact of this last effect is not obvious, and its resolution will require further study. 

In this paper we considered the effects of our merging algorithm in pure dijet analyses. 
Partially this was due to the availability of data; in addition it is worthwhile to first consider the effects of 
a new method in a simpler environment where there is no expectation of new physics.
Nevertheless it  is also important to apply this method to processes other than those which are purely QCD.
Since one of the primary motivations was to assess the impact of jet vetoes
that are relevant for Higgs plus dijets studies, this process is the next natural arena for study. 
We emphasise that this should not require any significant modifications to the method; 
the task is primarily an exercise in software development, rather than a theoretical challenge.

Finally, although we chose to implement this method for \pyt, in 
principle it should be possible to implement for other parton showers.  
It would be interesting to compare the effect of merging \HEJ with different choices of parton shower. 
It would also be informative to perform the jet analysis with a much harder
jet threshold, such that the sensitivity to the tunings of the
non-perturbative effects are reduced. This would result in a much cleaner
comparison of the perturbative merits.


Although we have been able to draw many positive conclusions by comparing with experimental data, the 
cuts that were chosen are not conducive to examining the effect of high energy logarithms. Both these points entail that it is difficult to discriminate 
between theoretical predictions that model different physics and should be expected to differ. 
We hope that as more data is collected, future analyses will examine a
similar set of observables but with more inclusive cuts, as discussed.

This work has reinforced the notion that the interplay between different types of logarithms is not necessarily straightforward, and that there are circumstances under which
the combination of two all-order summations is necessary. We hope that the merging algorithm we have developed may be used in future as a tool to inform 
analyses what selection of cuts and observables are sensitive to parton shower effects, high energy effects, or both. 



\acknowledgments

We would like to thank Stefan Prestel, Torbj\"orn Sj\"ostrand and Jennifer
M.~Smillie for discussions on the merging, analyses and early drafts. This
project has received funding from the European Union's Horizon 2020 research
and innovation programme under the Marie Sk\l{}odowska-Curie grant agreement
No 722104 and contract PITN-GA-2012-315877, the Swedish research council
(contracts 2016-03291 and 2016-05996) and the UK Science and Technology
Facilities Council (STFC).


\bibliographystyle{JHEP}  
\bibliography{refs} 

\providecommand{\href}[2]{#2}\begingroup\raggedright\begin{thebibliography}{10}

\bibitem{Abazov:2013gpa}
{\scshape D0} collaboration, V.~M. Abazov et~al., \emph{{Studies of W boson
  plus jets production in $p\bar{p}$ collisions at $\sqrt{s}=1.96$ TeV}},
  \href{http://dx.doi.org/10.1103/PhysRevD.88.092001}{\emph{Phys. Rev.} {\bf
  D88} (2013) 092001}, [\href{http://arxiv.org/abs/1302.6508}{{\tt
  1302.6508}}].

\bibitem{Aad:2011jz}
{\scshape ATLAS} collaboration, G.~Aad et~al., \emph{{Measurement of dijet
  production with a veto on additional central jet activity in $pp$ collisions
  at $\sqrt{s}=7$ TeV using the ATLAS detector}},
  \href{http://dx.doi.org/10.1007/JHEP09(2011)053}{\emph{JHEP} {\bf 09} (2011)
  053}, [\href{http://arxiv.org/abs/1107.1641}{{\tt 1107.1641}}].

\bibitem{Aad:2014pua}
{\scshape ATLAS} collaboration, G.~Aad et~al., \emph{{Measurements of jet
  vetoes and azimuthal decorrelations in dijet events produced in $pp$
  collisions at $\sqrt{s}=7\,\mathrm{TeV}$ using the ATLAS detector}},
  \href{http://dx.doi.org/10.1140/epjc/s10052-014-3117-7}{\emph{Eur. Phys. J.}
  {\bf C74} (2014) 3117}, [\href{http://arxiv.org/abs/1407.5756}{{\tt
  1407.5756}}].

\bibitem{Aad:2014qxa}
{\scshape ATLAS} collaboration, G.~Aad et~al., \emph{{Measurements of the W
  production cross sections in association with jets with the ATLAS detector}},
  \href{http://dx.doi.org/10.1140/epjc/s10052-015-3262-7}{\emph{Eur. Phys. J.}
  {\bf C75} (2015) 82}, [\href{http://arxiv.org/abs/1409.8639}{{\tt
  1409.8639}}].

\bibitem{Chatrchyan:2012gwa}
{\scshape CMS} collaboration, S.~Chatrchyan et~al., \emph{{Measurement of the
  inclusive production cross sections for forward jets and for dijet events
  with one forward and one central jet in $pp$ collisions at $\sqrt{s}=7$
  TeV}}, \href{http://dx.doi.org/10.1007/JHEP06(2012)036}{\emph{JHEP} {\bf 06}
  (2012) 036}, [\href{http://arxiv.org/abs/1202.0704}{{\tt 1202.0704}}].

\bibitem{Aad:2015nda}
{\scshape ATLAS} collaboration, G.~Aad et~al., \emph{{Measurement of four-jet
  differential cross sections in $\sqrt{s}=8$ TeV proton-proton collisions
  using the ATLAS detector}},
  \href{http://dx.doi.org/10.1007/JHEP12(2015)105}{\emph{JHEP} {\bf 12} (2015)
  105}, [\href{http://arxiv.org/abs/1509.07335}{{\tt 1509.07335}}].

\bibitem{Khachatryan:2016udy}
{\scshape CMS} collaboration, V.~Khachatryan et~al., \emph{{Azimuthal
  decorrelation of jets widely separated in rapidity in pp collisions at $
  \sqrt{s}=7 $ TeV}},
  \href{http://dx.doi.org/10.1007/JHEP08(2016)139}{\emph{JHEP} {\bf 08} (2016)
  139}, [\href{http://arxiv.org/abs/1601.06713}{{\tt 1601.06713}}].

\bibitem{Fadin:1975cb}
V.~S. Fadin, E.~A. Kuraev and L.~N. Lipatov, \emph{{On the Pomeranchuk
  Singularity in Asymptotically Free Theories}},
  \href{http://dx.doi.org/10.1016/0370-2693(75)90524-9}{\emph{Phys. Lett.} {\bf
  60B} (1975) 50--52}.

\bibitem{Kuraev:1976ge}
E.~A. Kuraev, L.~N. Lipatov and V.~S. Fadin, \emph{{Multi - Reggeon Processes
  in the Yang-Mills Theory}}, {\emph{Sov. Phys. JETP} {\bf 44} (1976)
  443--450}.

\bibitem{Kuraev:1977fs}
E.~A. Kuraev, L.~N. Lipatov and V.~S. Fadin, \emph{{The Pomeranchuk Singularity
  in Nonabelian Gauge Theories}}, {\emph{Sov. Phys. JETP} {\bf 45} (1977)
  199--204}.

\bibitem{Balitsky:1978ic}
I.~I. Balitsky and L.~N. Lipatov, \emph{{The Pomeranchuk Singularity in Quantum
  Chromodynamics}}, {\emph{Sov. J. Nucl. Phys.} {\bf 28} (1978) 822--829}.

\bibitem{Chatrchyan:2012pb}
{\scshape CMS} collaboration, S.~Chatrchyan et~al., \emph{{Ratios of dijet
  production cross sections as a function of the absolute difference in
  rapidity between jets in proton-proton collisions at $\sqrt{s}=7$ TeV}},
  \href{http://dx.doi.org/10.1140/epjc/s10052-012-2216-6}{\emph{Eur. Phys. J.}
  {\bf C72} (2012) 2216}, [\href{http://arxiv.org/abs/1204.0696}{{\tt
  1204.0696}}].

\bibitem{Chatrchyan:2013jya}
{\scshape CMS} collaboration, S.~Chatrchyan et~al., \emph{{Measurement of the
  hadronic activity in events with a Z and two jets and extraction of the cross
  section for the electroweak production of a Z with two jets in pp collisions
  at $\sqrt{s}$ = 7 TeV}},
  \href{http://dx.doi.org/10.1007/JHEP10(2013)062}{\emph{JHEP} {\bf 10} (2013)
  062}, [\href{http://arxiv.org/abs/1305.7389}{{\tt 1305.7389}}].

\bibitem{Aaboud:2017fye}
{\scshape ATLAS} collaboration, M.~Aaboud et~al., \emph{{Measurements of
  electroweak $Wjj$ production and constraints on anomalous gauge couplings
  with the ATLAS detector}},
  \href{http://dx.doi.org/10.1140/epjc/s10052-017-5007-2}{\emph{Eur. Phys. J.}
  {\bf C77} (2017) 474}, [\href{http://arxiv.org/abs/1703.04362}{{\tt
  1703.04362}}].

\bibitem{Dokshitzer:1977sg}
Y.~L. Dokshitzer, \emph{{Calculation of the Structure Functions for Deep
  Inelastic Scattering and e+ e- Annihilation by Perturbation Theory in Quantum
  Chromodynamics.}}, {\emph{Sov. Phys. JETP} {\bf 46} (1977) 641--653}.

\bibitem{Gribov:1972ri}
V.~N. Gribov and L.~N. Lipatov, \emph{{Deep inelastic e p scattering in
  perturbation theory}}, {\emph{Sov. J. Nucl. Phys.} {\bf 15} (1972) 438--450}.

\bibitem{Altarelli:1977zs}
G.~Altarelli and G.~Parisi, \emph{{Asymptotic Freedom in Parton Language}},
  \href{http://dx.doi.org/10.1016/0550-3213(77)90384-4}{\emph{Nucl. Phys.} {\bf
  B126} (1977) 298--318}.

\bibitem{Ciafaloni:1987ur}
M.~Ciafaloni, \emph{{Coherence Effects in Initial Jets at Small q**2 / s}},
  \href{http://dx.doi.org/10.1016/0550-3213(88)90380-X}{\emph{Nucl. Phys.} {\bf
  B296} (1988) 49--74}.

\bibitem{Catani:1989yc}
S.~Catani, F.~Fiorani and G.~Marchesini, \emph{{QCD Coherence in Initial State
  Radiation}},
  \href{http://dx.doi.org/10.1016/0370-2693(90)91938-8}{\emph{Phys. Lett.} {\bf
  B234} (1990) 339--345}.

\bibitem{Catani:1989sg}
S.~Catani, F.~Fiorani and G.~Marchesini, \emph{{Small x Behavior of Initial
  State Radiation in Perturbative QCD}},
  \href{http://dx.doi.org/10.1016/0550-3213(90)90342-B}{\emph{Nucl. Phys.} {\bf
  B336} (1990) 18--85}.

\bibitem{Marchesini:1994wr}
G.~Marchesini, \emph{{QCD coherence in the structure function and associated
  distributions at small x}},
  \href{http://dx.doi.org/10.1016/0550-3213(95)00149-M}{\emph{Nucl. Phys.} {\bf
  B445} (1995) 49--80}, [\href{http://arxiv.org/abs/hep-ph/9412327}{{\tt
  hep-ph/9412327}}].

\bibitem{Jung:2000hk}
H.~Jung and G.~P. Salam, \emph{{Hadronic final state predictions from CCFM: The
  Hadron level Monte Carlo generator CASCADE}},
  \href{http://dx.doi.org/10.1007/s100520100604}{\emph{Eur. Phys. J.} {\bf C19}
  (2001) 351--360}, [\href{http://arxiv.org/abs/hep-ph/0012143}{{\tt
  hep-ph/0012143}}].

\bibitem{Jung:2010si}
H.~Jung et~al., \emph{{The CCFM Monte Carlo generator CASCADE version 2.2.03}},
  \href{http://dx.doi.org/10.1140/epjc/s10052-010-1507-z}{\emph{Eur. Phys. J.}
  {\bf C70} (2010) 1237--1249}, [\href{http://arxiv.org/abs/1008.0152}{{\tt
  1008.0152}}].

\bibitem{Sjostrand:2006za}
T.~Sj{\"o}strand, S.~Mrenna and P.~Z. Skands, \emph{{PYTHIA 6.4 Physics and
  Manual}}, \href{http://dx.doi.org/10.1088/1126-6708/2006/05/026}{\emph{JHEP}
  {\bf 05} (2006) 026}, [\href{http://arxiv.org/abs/hep-ph/0603175}{{\tt
  hep-ph/0603175}}].

\bibitem{Mueller:1986ey}
A.~H. Mueller and H.~Navelet, \emph{{An Inclusive Minijet Cross-Section and the
  Bare Pomeron in QCD}},
  \href{http://dx.doi.org/10.1016/0550-3213(87)90705-X}{\emph{Nucl. Phys.} {\bf
  B282} (1987) 727--744}.

\bibitem{Orr:1997im}
L.~H. Orr and W.~J. Stirling, \emph{{Dijet production at hadron hadron
  colliders in the BFKL approach}},
  \href{http://dx.doi.org/10.1103/PhysRevD.56.5875}{\emph{Phys. Rev.} {\bf D56}
  (1997) 5875--5884}, [\href{http://arxiv.org/abs/hep-ph/9706529}{{\tt
  hep-ph/9706529}}].

\bibitem{Orr:1998ps}
L.~H. Orr and W.~J. Stirling, \emph{{BFKL physics in dijet production at the
  LHC}}, \href{http://dx.doi.org/10.1016/S0370-2693(98)00864-8}{\emph{Phys.
  Lett.} {\bf B436} (1998) 372--378},
  [\href{http://arxiv.org/abs/hep-ph/9806371}{{\tt hep-ph/9806371}}].

\bibitem{Andersen:2009nu}
J.~R. Andersen and J.~M. Smillie, \emph{{Constructing All-Order Corrections to
  Multi-Jet Rates}},
  \href{http://dx.doi.org/10.1007/JHEP01(2010)039}{\emph{JHEP} {\bf 01} (2010)
  039}, [\href{http://arxiv.org/abs/0908.2786}{{\tt 0908.2786}}].

\bibitem{Andersen:2009he}
J.~R. Andersen and J.~M. Smillie, \emph{{The Factorisation of the t-channel
  Pole in Quark-Gluon Scattering}},
  \href{http://dx.doi.org/10.1103/PhysRevD.81.114021}{\emph{Phys. Rev.} {\bf
  D81} (2010) 114021}, [\href{http://arxiv.org/abs/0910.5113}{{\tt
  0910.5113}}].

\bibitem{Andersen:2011hs}
J.~R. Andersen and J.~M. Smillie, \emph{{Multiple Jets at the LHC with High
  Energy Jets}}, \href{http://dx.doi.org/10.1007/JHEP06(2011)010}{\emph{JHEP}
  {\bf 06} (2011) 010}, [\href{http://arxiv.org/abs/1101.5394}{{\tt
  1101.5394}}].

\bibitem{DelDuca:2001fn}
V.~Del~Duca, W.~Kilgore, C.~Oleari, C.~Schmidt and D.~Zeppenfeld, \emph{{Gluon
  fusion contributions to H + 2 jet production}},
  \href{http://dx.doi.org/10.1016/S0550-3213(01)00446-1}{\emph{Nucl. Phys.}
  {\bf B616} (2001) 367--399}, [\href{http://arxiv.org/abs/hep-ph/0108030}{{\tt
  hep-ph/0108030}}].

\bibitem{Klamke:2007cu}
G.~Klamke and D.~Zeppenfeld, \emph{{Higgs plus two jet production via gluon
  fusion as a signal at the CERN LHC}},
  \href{http://dx.doi.org/10.1088/1126-6708/2007/04/052}{\emph{JHEP} {\bf 04}
  (2007) 052}, [\href{http://arxiv.org/abs/hep-ph/0703202}{{\tt
  hep-ph/0703202}}].

\bibitem{Andersen:2010zx}
J.~R. Andersen, K.~Arnold and D.~Zeppenfeld, \emph{{Azimuthal Angle
  Correlations for Higgs Boson plus Multi-Jet Events}},
  \href{http://dx.doi.org/10.1007/JHEP06(2010)091}{\emph{JHEP} {\bf 06} (2010)
  091}, [\href{http://arxiv.org/abs/1001.3822}{{\tt 1001.3822}}].

\bibitem{Lonnblad:1992tz}
L.~L{\"o}nnblad, \emph{{ARIADNE version 4: A Program for simulation of QCD
  cascades implementing the color dipole model}},
  \href{http://dx.doi.org/10.1016/0010-4655(92)90068-A}{\emph{Comput. Phys.
  Commun.} {\bf 71} (1992) 15--31}.

\bibitem{Andersen:2011zd}
J.~R. Andersen, L.~L{\"o}nnblad and J.~M. Smillie, \emph{{A Parton Shower for
  High Energy Jets}},
  \href{http://dx.doi.org/10.1007/JHEP07(2011)110}{\emph{JHEP} {\bf 07} (2011)
  110}, [\href{http://arxiv.org/abs/1104.1316}{{\tt 1104.1316}}].

\bibitem{Sjostrand:2004ef}
T.~Sj{\"o}strand and P.~Z. Skands, \emph{{Transverse-momentum-ordered showers
  and interleaved multiple interactions}},
  \href{http://dx.doi.org/10.1140/epjc/s2004-02084-y}{\emph{Eur. Phys. J.} {\bf
  C39} (2005) 129--154}, [\href{http://arxiv.org/abs/hep-ph/0408302}{{\tt
  hep-ph/0408302}}].

\bibitem{Sjostrand:2014zea}
T.~Sj{\"o}strand, S.~Ask, J.~R. Christiansen, R.~Corke, N.~Desai, P.~Ilten
  et~al., \emph{{An Introduction to PYTHIA 8.2}},
  \href{http://dx.doi.org/10.1016/j.cpc.2015.01.024}{\emph{Comput. Phys.
  Commun.} {\bf 191} (2015) 159--177},
  [\href{http://arxiv.org/abs/1410.3012}{{\tt 1410.3012}}].

\bibitem{Andersen:2017kfc}
J.~R. Andersen, T.~Hapola, A.~Maier and J.~M. Smillie, \emph{{Higgs Boson Plus
  Dijets: Higher Order Corrections}},
  \href{http://arxiv.org/abs/1706.01002}{{\tt 1706.01002}}.

\bibitem{Alwall:2014hca}
J.~Alwall, R.~Frederix, S.~Frixione, V.~Hirschi, F.~Maltoni, O.~Mattelaer
  et~al., \emph{{The automated computation of tree-level and next-to-leading
  order differential cross sections, and their matching to parton shower
  simulations}}, \href{http://dx.doi.org/10.1007/JHEP07(2014)079}{\emph{JHEP}
  {\bf 07} (2014) 079}, [\href{http://arxiv.org/abs/1405.0301}{{\tt
  1405.0301}}].

\bibitem{Parke:1986gb}
S.~J. Parke and T.~R. Taylor, \emph{{An Amplitude for $n$ Gluon Scattering}},
  \href{http://dx.doi.org/10.1103/PhysRevLett.56.2459}{\emph{Phys. Rev. Lett.}
  {\bf 56} (1986) 2459}.

\bibitem{DelDuca:1993pp}
V.~Del~Duca, \emph{{Parke-Taylor amplitudes in the multi - Regge kinematics}},
  \href{http://dx.doi.org/10.1103/PhysRevD.48.5133}{\emph{Phys. Rev.} {\bf D48}
  (1993) 5133--5139}, [\href{http://arxiv.org/abs/hep-ph/9304259}{{\tt
  hep-ph/9304259}}].

\bibitem{DelDuca:1995zy}
V.~Del~Duca, \emph{{Equivalence of the Parke-Taylor and the
  Fadin-Kuraev-Lipatov amplitudes in the high-energy limit}},
  \href{http://dx.doi.org/10.1103/PhysRevD.52.1527}{\emph{Phys. Rev.} {\bf D52}
  (1995) 1527--1534}, [\href{http://arxiv.org/abs/hep-ph/9503340}{{\tt
  hep-ph/9503340}}].

\bibitem{Boos:2001cv}
E.~Boos et~al., \emph{{Generic user process interface for event generators}},
  in \emph{{Physics at TeV colliders. Proceedings, Euro Summer School, Les
  Houches, France, May 21-June 1, 2001}}, 2001.
\newblock \href{http://arxiv.org/abs/hep-ph/0109068}{{\tt hep-ph/0109068}}.

\bibitem{Andersson:1988gp}
B.~Andersson, G.~Gustafson, L.~L{\"o}nnblad and U.~Pettersson, \emph{{Coherence
  Effects in Deep Inelastic Scattering}},
  \href{http://dx.doi.org/10.1007/BF01550942}{\emph{Z. Phys.} {\bf C43} (1989)
  625}.

\bibitem{Sjostrand:1987su}
T.~Sj{\"o}strand and M.~van Zijl, \emph{{A Multiple Interaction Model for the
  Event Structure in Hadron Collisions}},
  \href{http://dx.doi.org/10.1103/PhysRevD.36.2019}{\emph{Phys. Rev.} {\bf D36}
  (1987) 2019}.

\bibitem{Lonnblad:2001iq}
L.~Lonnblad, \emph{{Correcting the color dipole cascade model with fixed order
  matrix elements}},
  \href{http://dx.doi.org/10.1088/1126-6708/2002/05/046}{\emph{JHEP} {\bf 05}
  (2002) 046}, [\href{http://arxiv.org/abs/hep-ph/0112284}{{\tt
  hep-ph/0112284}}].

\bibitem{Lavesson:2005xu}
N.~Lavesson and L.~Lonnblad, \emph{{W+jets matrix elements and the dipole
  cascade}}, \href{http://dx.doi.org/10.1088/1126-6708/2005/07/054}{\emph{JHEP}
  {\bf 07} (2005) 054}, [\href{http://arxiv.org/abs/hep-ph/0503293}{{\tt
  hep-ph/0503293}}].

\bibitem{Lonnblad:2011xx}
L.~Lonnblad and S.~Prestel, \emph{{Matching Tree-Level Matrix Elements with
  Interleaved Showers}},
  \href{http://dx.doi.org/10.1007/JHEP03(2012)019}{\emph{JHEP} {\bf 03} (2012)
  019}, [\href{http://arxiv.org/abs/1109.4829}{{\tt 1109.4829}}].

\bibitem{Catani:2001cc}
S.~Catani, F.~Krauss, R.~Kuhn and B.~R. Webber, \emph{{QCD matrix elements +
  parton showers}},
  \href{http://dx.doi.org/10.1088/1126-6708/2001/11/063}{\emph{JHEP} {\bf 11}
  (2001) 063}, [\href{http://arxiv.org/abs/hep-ph/0109231}{{\tt
  hep-ph/0109231}}].

\bibitem{Mangano:2006rw}
M.~L. Mangano, M.~Moretti, F.~Piccinini and M.~Treccani, \emph{{Matching matrix
  elements and shower evolution for top-quark production in hadronic
  collisions}},
  \href{http://dx.doi.org/10.1088/1126-6708/2007/01/013}{\emph{JHEP} {\bf 01}
  (2007) 013}, [\href{http://arxiv.org/abs/hep-ph/0611129}{{\tt
  hep-ph/0611129}}].

\bibitem{Bengtsson:1987rw}
M.~Bengtsson and T.~Sjostrand, \emph{{Parton Showers in Leptoproduction
  Events}}, \href{http://dx.doi.org/10.1007/BF01578142}{\emph{Z. Phys.} {\bf
  C37} (1988) 465}.

\bibitem{Ellis:1991qj}
R.~K. Ellis, W.~J. Stirling and B.~R. Webber, \emph{{QCD and Collider
  Physics}}, vol.~8 of \emph{Camb. Monogr. Part. Phys. Nucl. Phys. Cosmol.}
\newblock {Cambridge University Press}, 1996.

\bibitem{Alioli:2012tp}
S.~Alioli, J.~R. Andersen, C.~Oleari, E.~Re and J.~M. Smillie, \emph{{Probing
  higher-order corrections in dijet production at the LHC}},
  \href{http://dx.doi.org/10.1103/PhysRevD.85.114034}{\emph{Phys. Rev.} {\bf
  D85} (2012) 114034}, [\href{http://arxiv.org/abs/1202.1475}{{\tt
  1202.1475}}].

\bibitem{Skands:2014pea}
P.~Skands, S.~Carrazza and J.~Rojo, \emph{{Tuning PYTHIA 8.1: the Monash 2013
  Tune}}, \href{http://dx.doi.org/10.1140/epjc/s10052-014-3024-y}{\emph{Eur.
  Phys. J.} {\bf C74} (2014) 3024}, [\href{http://arxiv.org/abs/1404.5630}{{\tt
  1404.5630}}].

\bibitem{Aad:2011kq}
{\scshape ATLAS} collaboration, G.~Aad et~al., \emph{{Study of Jet Shapes in
  Inclusive Jet Production in $pp$ Collisions at $\sqrt{s}=7$ TeV using the
  ATLAS Detector}},
  \href{http://dx.doi.org/10.1103/PhysRevD.83.052003}{\emph{Phys. Rev.} {\bf
  D83} (2011) 052003}, [\href{http://arxiv.org/abs/1101.0070}{{\tt
  1101.0070}}].

\bibitem{Cacciari:2008gp}
M.~Cacciari, G.~P. Salam and G.~Soyez, \emph{{The Anti-k(t) jet clustering
  algorithm}},
  \href{http://dx.doi.org/10.1088/1126-6708/2008/04/063}{\emph{JHEP} {\bf 04}
  (2008) 063}, [\href{http://arxiv.org/abs/0802.1189}{{\tt 0802.1189}}].

\bibitem{Buckley:2010ar}
A.~Buckley, J.~Butterworth, L.~Lonnblad, D.~Grellscheid, H.~Hoeth, J.~Monk
  et~al., \emph{{Rivet user manual}},
  \href{http://dx.doi.org/10.1016/j.cpc.2013.05.021}{\emph{Comput. Phys.
  Commun.} {\bf 184} (2013) 2803--2819},
  [\href{http://arxiv.org/abs/1003.0694}{{\tt 1003.0694}}].

\bibitem{Dulat:2015mca}
S.~Dulat, T.-J. Hou, J.~Gao, M.~Guzzi, J.~Huston, P.~Nadolsky et~al.,
  \emph{{New parton distribution functions from a global analysis of quantum
  chromodynamics}},
  \href{http://dx.doi.org/10.1103/PhysRevD.93.033006}{\emph{Phys. Rev.} {\bf
  D93} (2016) 033006}, [\href{http://arxiv.org/abs/1506.07443}{{\tt
  1506.07443}}].

\bibitem{Buckley:2014ana}
A.~Buckley, J.~Ferrando, S.~Lloyd, K.~Nordström, B.~Page, M.~Rüfenacht
  et~al., \emph{{LHAPDF6: parton density access in the LHC precision era}},
  \href{http://dx.doi.org/10.1140/epjc/s10052-015-3318-8}{\emph{Eur. Phys. J.}
  {\bf C75} (2015) 132}, [\href{http://arxiv.org/abs/1412.7420}{{\tt
  1412.7420}}].

\bibitem{Frixione:1997ks}
S.~Frixione and G.~Ridolfi, \emph{{Jet photoproduction at HERA}},
  \href{http://dx.doi.org/10.1016/S0550-3213(97)00575-0}{\emph{Nucl. Phys.}
  {\bf B507} (1997) 315--333}, [\href{http://arxiv.org/abs/hep-ph/9707345}{{\tt
  hep-ph/9707345}}].

\bibitem{Nason:2004rx}
P.~Nason, \emph{{A New method for combining NLO QCD with shower Monte Carlo
  algorithms}},
  \href{http://dx.doi.org/10.1088/1126-6708/2004/11/040}{\emph{JHEP} {\bf 11}
  (2004) 040}, [\href{http://arxiv.org/abs/hep-ph/0409146}{{\tt
  hep-ph/0409146}}].

\bibitem{Frixione:2007vw}
S.~Frixione, P.~Nason and C.~Oleari, \emph{{Matching NLO QCD computations with
  Parton Shower simulations: the POWHEG method}},
  \href{http://dx.doi.org/10.1088/1126-6708/2007/11/070}{\emph{JHEP} {\bf 11}
  (2007) 070}, [\href{http://arxiv.org/abs/0709.2092}{{\tt 0709.2092}}].

\bibitem{Alioli:2010xa}
S.~Alioli, K.~Hamilton, P.~Nason, C.~Oleari and E.~Re, \emph{{Jet pair
  production in POWHEG}},
  \href{http://dx.doi.org/10.1007/JHEP04(2011)081}{\emph{JHEP} {\bf 04} (2011)
  081}, [\href{http://arxiv.org/abs/1012.3380}{{\tt 1012.3380}}].

\end{thebibliography}\endgroup

\end{document}